\documentclass[11pt,twoside]{article}
\usepackage{asp2004}
\usepackage{psfig}
\usepackage{epsf}
\usepackage{graphics}
\usepackage{lscape}
\def\edcomment#1{\iffalse\marginpar{\raggedright\sl#1\/}\else\relax\fi}
\reversemarginpar

\begin{document}
\title{Light Elements in Main Sequence Stars: Li, Be, B, C, O}
\author{Ann Merchant Boesgaard}
\affil{Institute for Astronomy, University of Hawai`i, Honolulu, HI 96822,
U.S.A.}

\begin{abstract}
The abundances of the {\it rare} light elements, Li, Be, and B, provide clues
about stellar structure and evolution, about Galactic evolution and about
their nucleosynthesis, including production during the Big Bang.  The
abundances of the {\it abundant} light elements, C and O, reveal information
about the chemical history of the Galaxy and the mass spectrum of early
generations of stars.
\end{abstract}

\thispagestyle{plain}

\section{The Rare Light Elements}

The trio of light elements, Li, Be, and B, offer a special opportunity to
discern the structure and processes occurring beneath the surfaces of stars.
The three are destroyed by fusion reactions at a few million K.  The isotope
of $^7$Li fuses with a proton at $T$ $\sim$2.5 x 10$^6$ K and higher which
corresponds to the inner 97.5\% (by mass) of the solar model.  For $^9$Be the
temperature is $\sim$3.5 x 10$^6$ K and higher, or 95\% (by mass).  The
relevant figures for the B isotopes ($^{10}$B and $^{11}$B) are near 5 x
10$^6$ K and 18-20\% (by mass).  Thus the surface zones in which each element
is preserved have different dimensions; the amount of each element remaining
on the stellar surface indicates how deep the mixing has been.  And since the
depletion in F and G dwarfs is apparently due to slow mixing, the depletion is
a function of the age of a star.

Due to the low abundances of these rare light elements in stellar atmospheres,
they are primarily observed in their respective resonance lines.  For F and G
stars these are 6707.74 and 6707.89 \AA\ of Li I, 3130.42 and 3131.06 \AA\
of Be II, and 2496.77 \AA\ of B I.  The Li I resonance doublet occurs in a
relatively clean part of the spectrum, but both the Be II and B I features are
in spectral regions crowded with other lines.  Extracting the abundances
requires the use of spectrum synthesis methods.  Examples of synthesized
spectra of all three elements can be found in Boesgaard et al.~(2004c).

\subsection{The Li and Be Dips in mid-F Dwarfs}

The discovery of the Li dip in the Hyades by Boesgaard \& Tripicco (1986) was
followed by a series of papers searching for such a dip in other open clusters
(e.g. Pilachowski, Booth \& Hobbs (1987), Boesgaard, Budge \& Ramsay (1988),
Hobbs \& Pilachowski (1988), Soderblom et al.~(1993), and more recently,
Steinhauer (2003).)  A search for a Be dip in the Hyades was done by Boesgaard
\& Budge (1989) with the tantalizing hint of such a dip.  It has been possible
with the Keck telescope and HIRES (Vogt et al.~1994) to re-examine the Hyades
and to extend the search for a Be dip to other clusters.  One cautionary note,
however, is that many of the young stars with mid-F spectral types (near the
center of the dip) are rotating sufficiently to affect the reliability of the
Be abundance determination.  Therefore we observed stars with $v$ sin $i$ $<$
20 km s$^{-1}$.

Boesgaard \& King (2002) found compelling evidence of a Be dip in the Hyades
which is $\sim$700 Myr old; this is shown in the left panel of Figure 1.  The
comparison of the Li and Be abundance is shown in the right panel of Figure 1.
The abundances are on the same scale and are normalized to the meteoritic
values of A(Li) = 3.30 and A(Be) = 1.42 (Grevesse \& Sauval 1998).  (We define
A(element) = log N(element)/N(H) + 12.00.)  There are two major differences in
the abundance-temperature profile between the two elements.  1) The Be dip in
the mid-F stars is not as deep as the Li dip.  2) There is no apparent
depletion of Be in the G stars in spite of the large (a factor of 100)
depletion of Li.  These results are not unexpected as the Be atoms need to be
mixed down deeper to higher temperatures within the star to be destroyed.  The
volume of the region where Be is preserved is larger than that where Li is
preserved indicating a larger number of surviving Be atoms initially.

\begin{figure}[!ht]
\plottwo{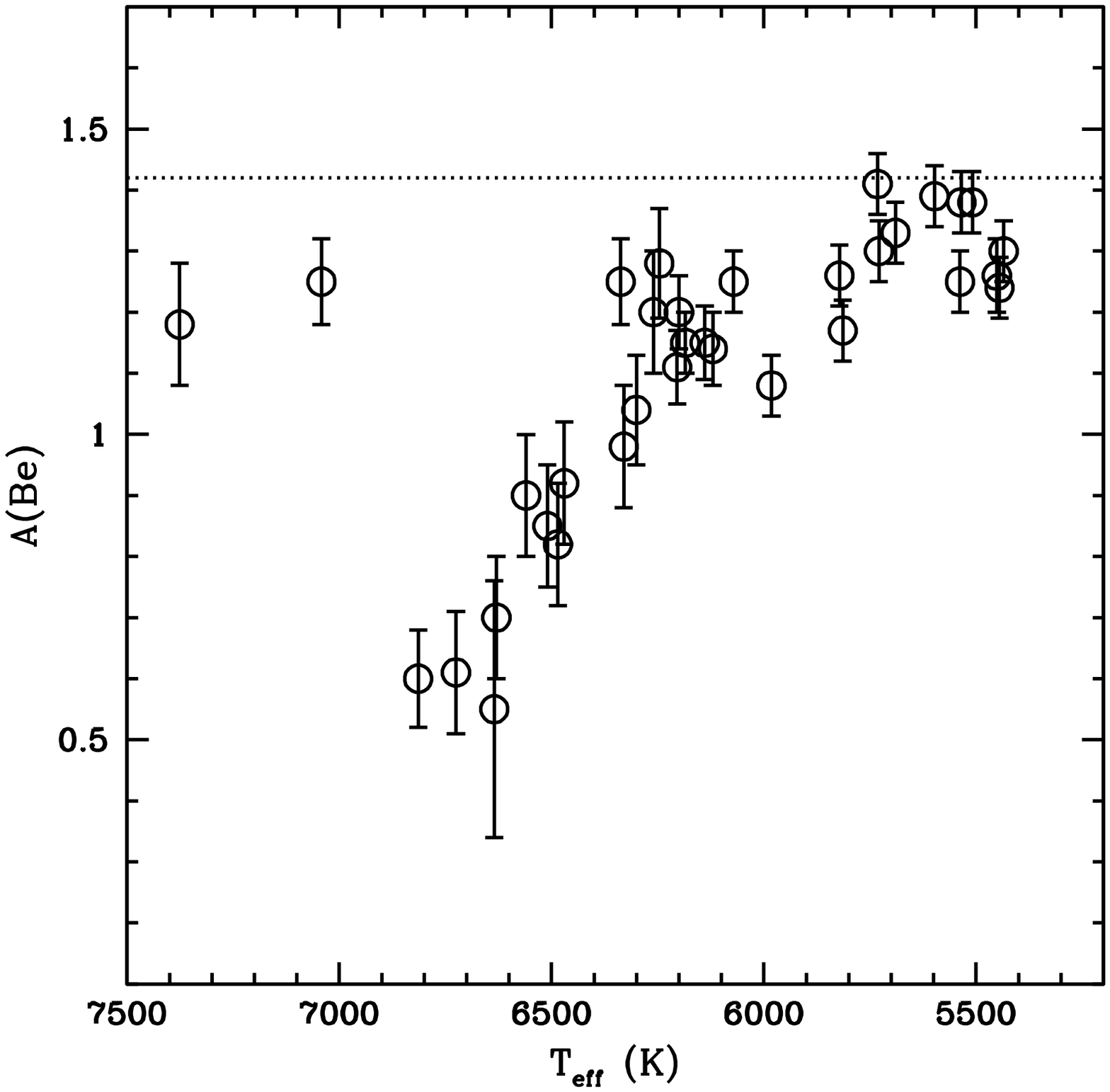}{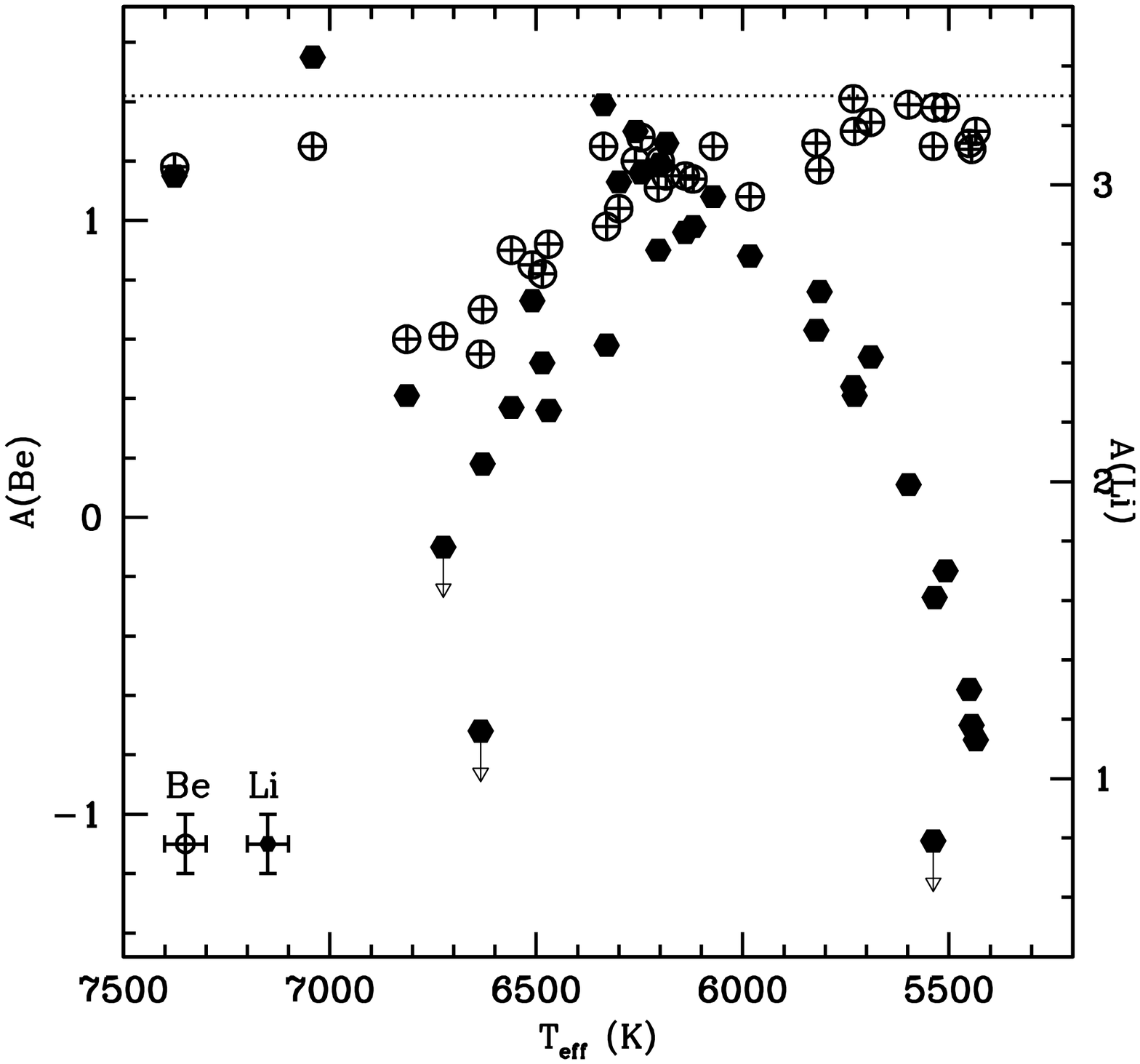}
\caption{Left: The Be dip in the Hyades.  Abundances of Be are shown as a
function of temperature.  The horizontal line corresponds to A(Be) in
meteorites of 1.42.  Right: A display of both Li and Be in the Hyades on the
same scale with A(Be) on the left y-axis and A(Li) on the right y-axis.  The
Li results are shown as hexagons and the Be results as encircled plus signs.
The two elements are normalized to their respective meteoritic abundances of
1.42 for Be and 3.30 for Li, shown by the horizontal line.  ApJ, 565, 587.}
\end{figure}

\begin{figure}[!ht]
\plottwo{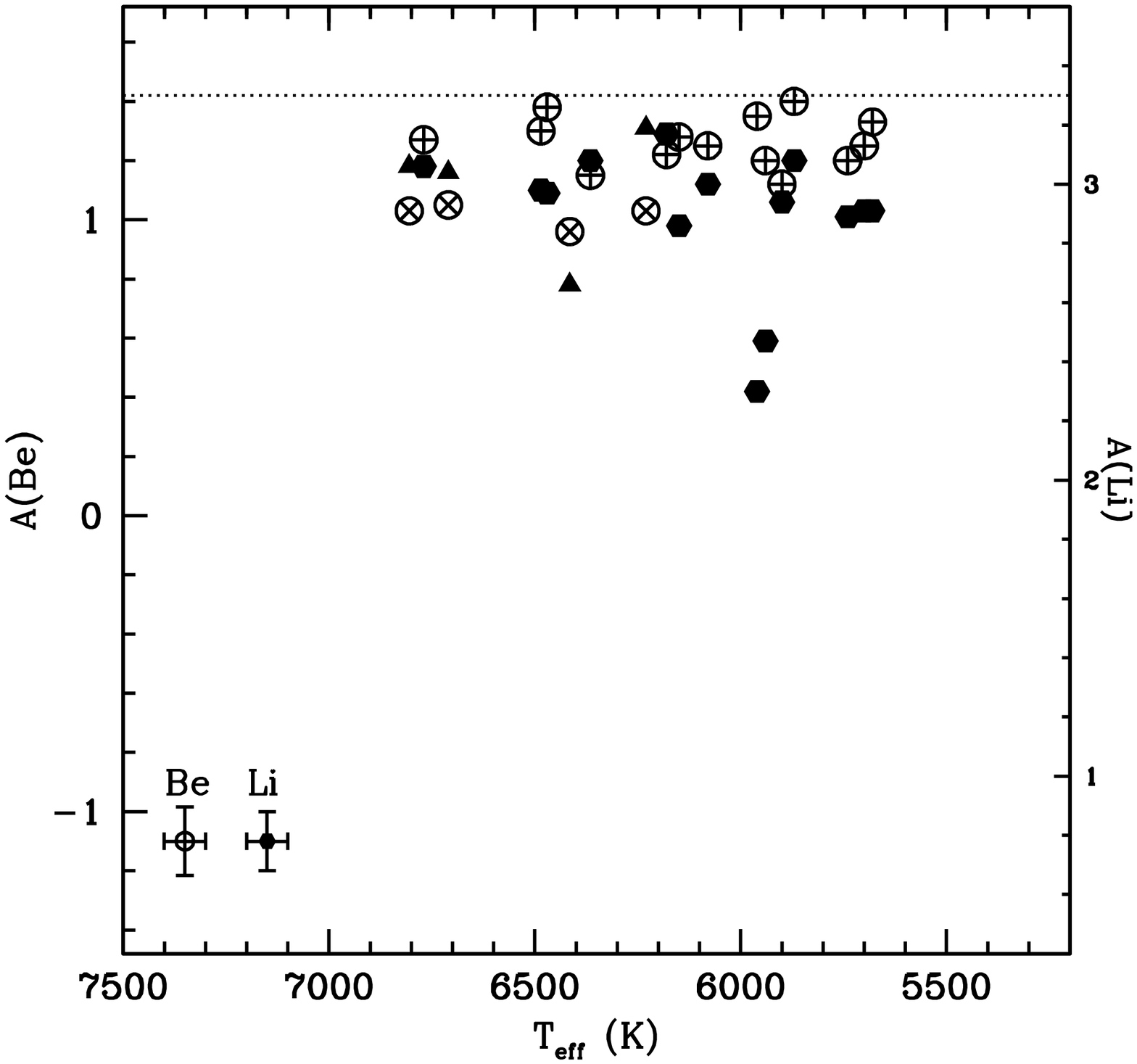}{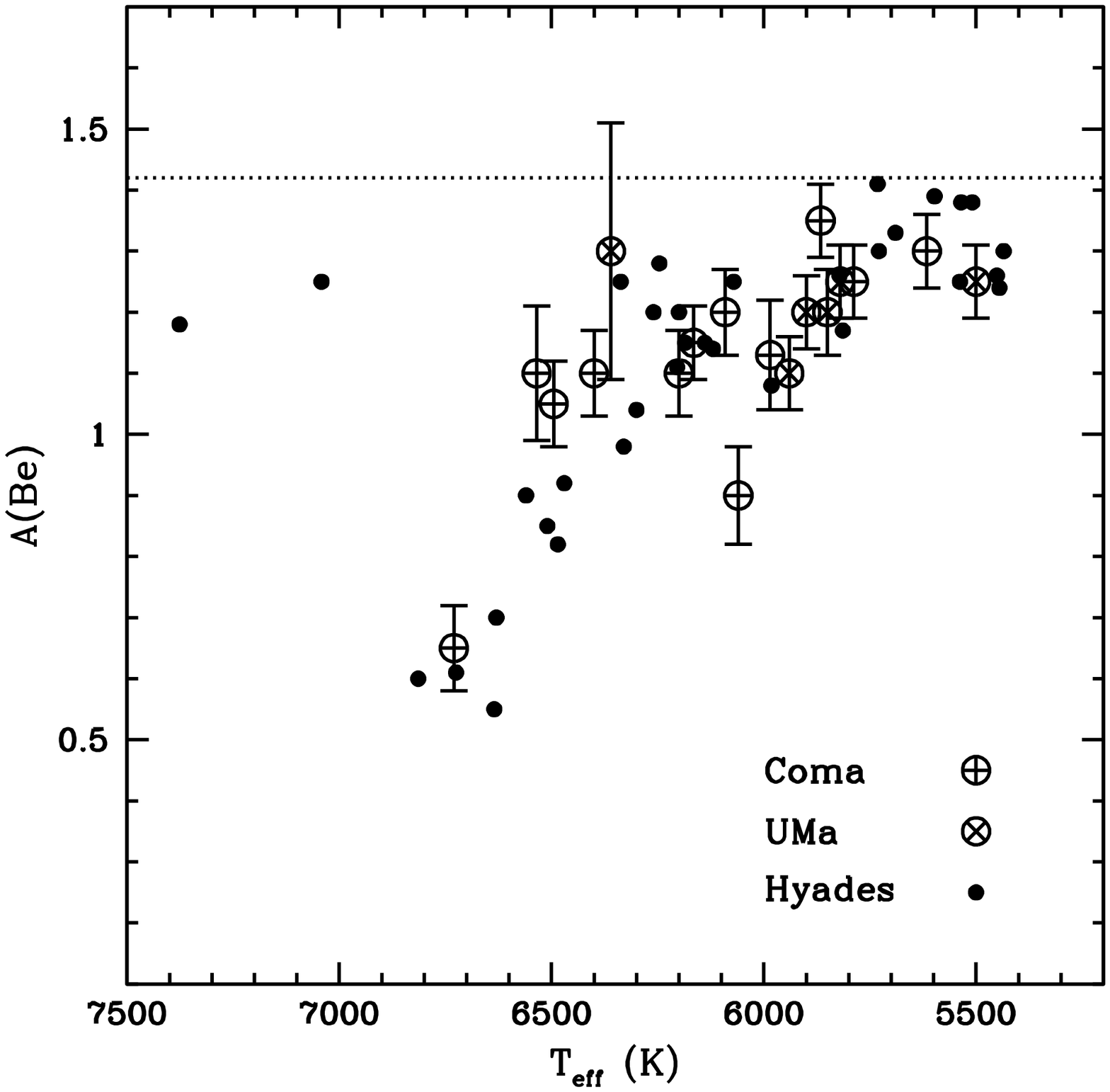}
\caption{Left: Li and Be abundances in the Pleiades and $\alpha$ Per samples.
The Pleiades Li results are shown as hexagons and the Be results as encircled
plus signs and the $\alpha$ Per results as triangle for Li and circled x signs
for Be.  ApJ 582, 410.  Right: Abundances of Be in the Coma and UMa clusters
compared with the Hyades (small dots).  The Coma results are shown as
encircled plus signs and Uma as circled x signs.  ApJ, 583, 955.}
\end{figure}

Other young and intermediate age clusters have been investigated for the
presence of a Be dip and to check Be in the G dwarfs.  Boesgaard, Armengaud \&
King (2003a) looked at the younger Pleiades and $\alpha$ Per clusters.  The
left panel of Figure 2 shows the scaled Li and Be abundances in those two
clusters.  There is no evidence of a Be dip in these clusters which are
$\sim$50 - 70 Myr old.  This indicates that the Li-Be dip is a phenomenon that
occurs on the main sequence, not during pre-main sequence evolution, after an
age of 100 Myr or so.

Other young clusters have been studied for Be including Coma and the UMa
moving group which are slightly younger than the Hyades at 300 - 500 Myr.
Boesgaard, Armengaud \& King (2003b) found evidence of Be dip in the Coma
cluster which can be seen in the right panel of Figure 2.  It is shown in
comparison with the Hyades Be dip and is on the same scale as the left panel
of Figure 1.  The pattern is similar to the Hyades with perhaps more
(intrinsic?) spread in A(Be).  The Praesepe cluster was also studied by
Boesgaard, Armengaud \& King (2004a).  They found depleted Be in the Li dip
region in Praesepe stars and summarized the results for five young to
intermediate age clusters.  The correlation between Li and Be in the cluster
stars was presented for stars in the temperature range from 5900 - 6650 K.
(This is discussed below in $\S$1.3.)

\subsection{Beryllium in Field Stars}

In addition to the Be studies in cluster stars Boesgaard et al.~(2004b) have
determined Li and Be in an array of F and G field dwarfs with high S/N spectra
primarily from Keck/HIRES (Be) and the UH 2.2-m coud\'e spectrograph (Li).
They have compiled the results from that and other studies, including the
cluster work, into plots showing the range and trends of A(Be) with both
effective temperature and metallicity.  Those results are shown in Figure 3.

It can be seen that there are stars at or near the meteoritic abundance at all
temperatures (left), but only at the higher metallicity values, near solar
[Fe/H] (right).  The field stars can show large Be depletions (two orders of
magnitude and more), but the (young) cluster stars in the Be dip region have
depletions of no more than a factor of 10.  For the cooler stars ($T$ $<$ 5700
K) field stars show depletions of up to a factor of 4, but the relatively
younger cluster stars have little or no Be depletion.  The field stars in the
Be dip may have no Be depletion while others have depletions larger than a
factor of 100.  The highly Be-depleted stars may be the older ones or those
with higher initial angular momentum.

The right panel of Figure 3 shows that the full range of A(Be) is seen at
solar metallicity, [Fe/H] = 0.0 $\pm$ 0.2.  The full range extends out to
stars with metallicities a factor of two less than solar ([Fe/H] = $-$0.3).
It appears that there is an upper envelope of A(Be) as a function of [Fe/H];
that is, the initial amount of Be in stars is correlated with its [Fe/H].
This appears to be the case for halo stars as well (e.g.~Duncan et al.~1998).

\begin{figure}[!ht]
\plottwo{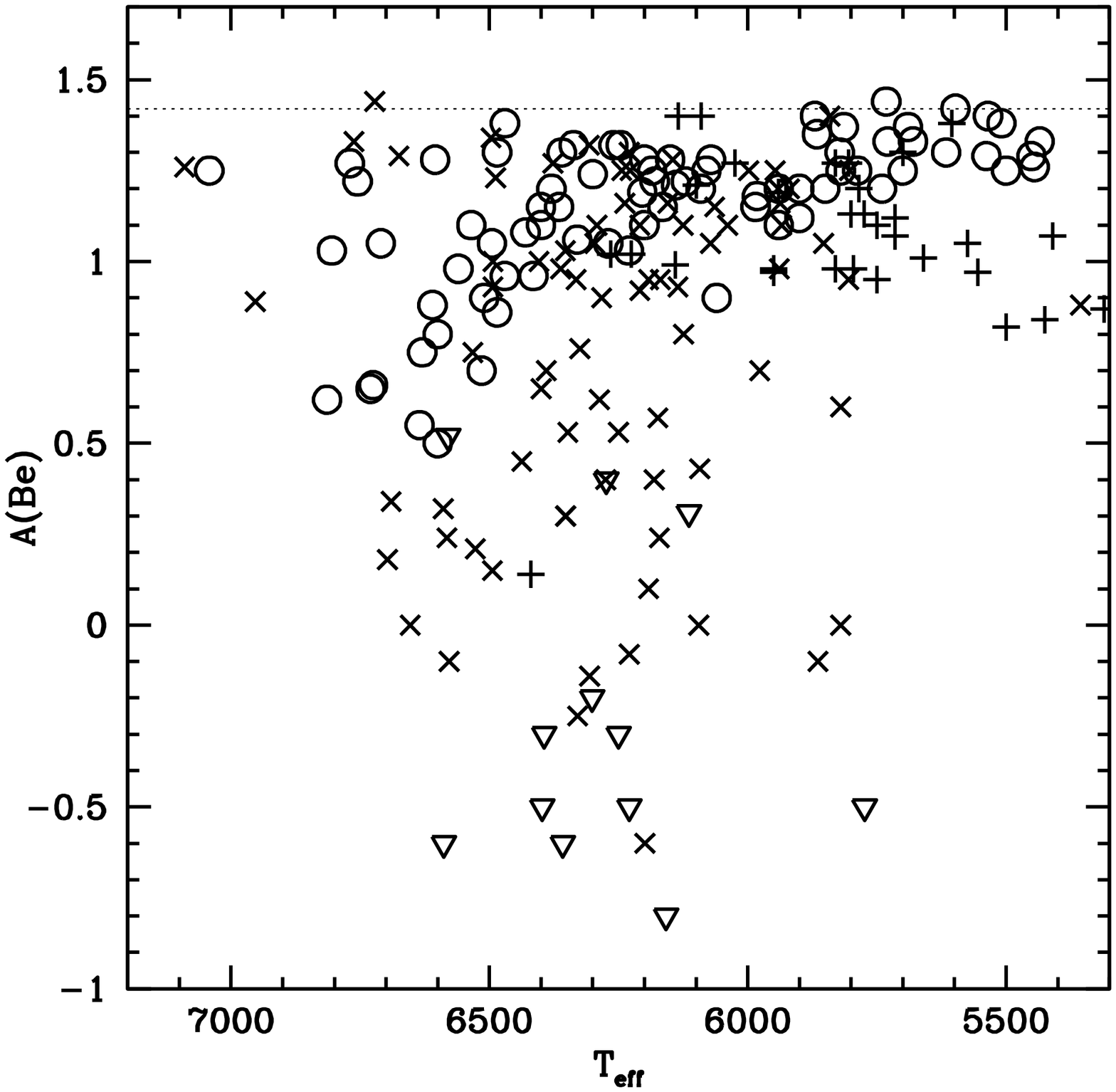}{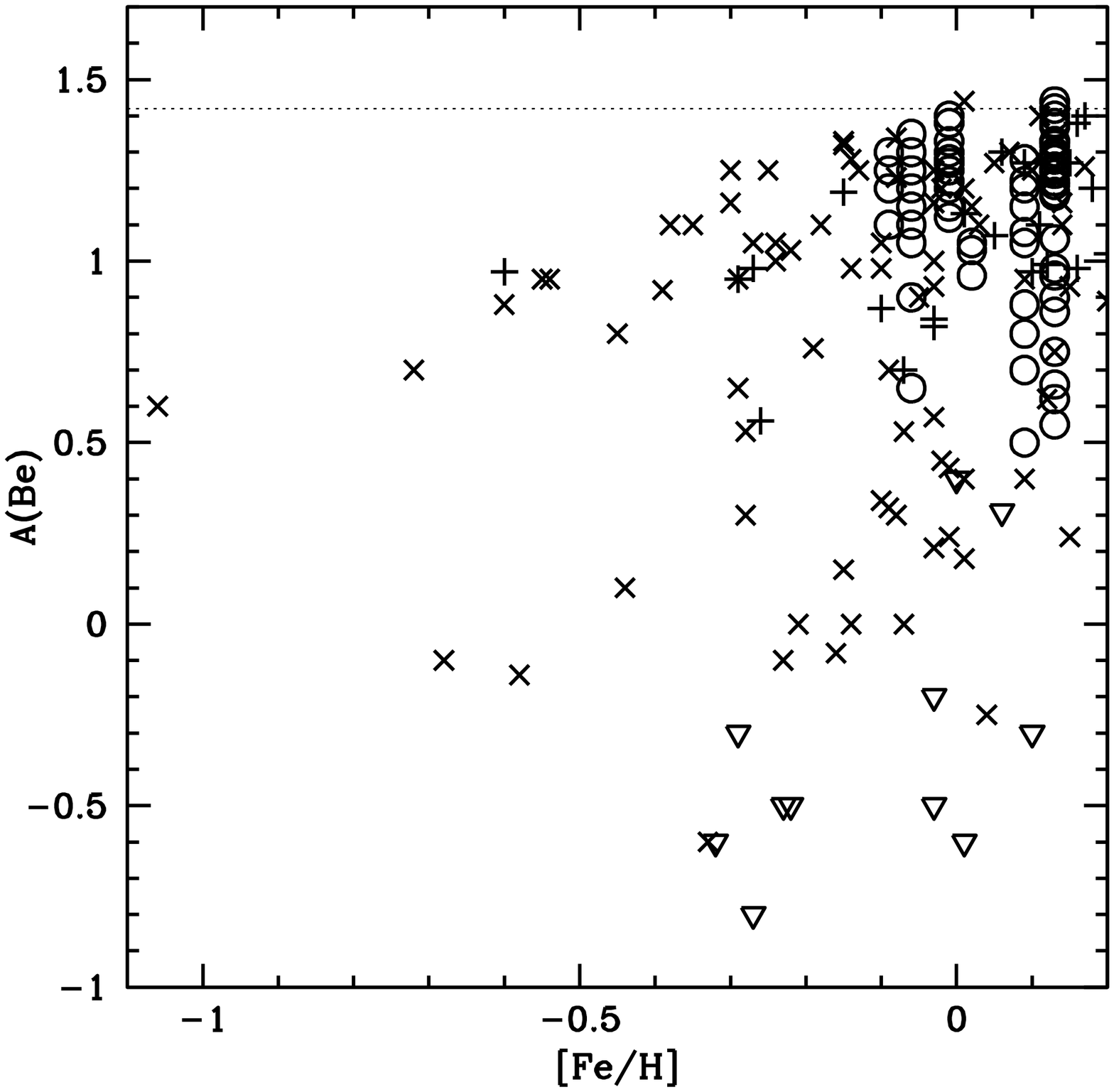}
\caption{Left: Values of A(Be) as a function of temperature.  This compilation
is from several sources; the crosses are field stars from Boesgaard et
al.~(2004b), Deliyannis et al.~(1998) and Boesgaard et al.~(2001).  The
inverted triangles represent upper limits on A(Be) from those papers.  The
open circles are cluster stars from Hyades (Boesgaard \& King 2002, Boesgaard
et al.~2004a) the Pleiades and $\alpha$ Per (Boesgaard, Armengaud \& King
2003a), Coma and UMa (Boesgaard, Armengaud \& King 2003b) and Praesepe
(Boesgaard, Armengaud \& King 2004a).  The plus signs are the stars with
exoplanets from Santos et al. (2002).  Right: Values of A(Be) in F and G
dwarfs as a function of [Fe/H].  The symbols are the same as in the left
panel.  ApJ, 613, 1202.}
\end{figure}

\subsection{The Correlation of Li and Be}

The cluster and field star data on Li and Be can be assembled to investigate
the correlation between the abundances of these elements.  Boesgaard et
al.~(2004b) have done this and have found correlations for the 88 field and
cluster stars between 5900 - 6650 K.  They have divided this into two
temperatures ranges corresponding to the cool side of the Li-Be dip, 6300 -
6650 K, where Li and Be are increasing as the temperature decreases and the
cooler group of early G stars near the Hyades Li peak, 5900 - 6300 K.  The
correlations for these two groups are shown in Figure 4, along with the best
fit slope for each group.  A light line with the slope for the other group is
also shown to illustrate differences between the two temperature groups.

\begin{figure}[!ht]
\plottwo{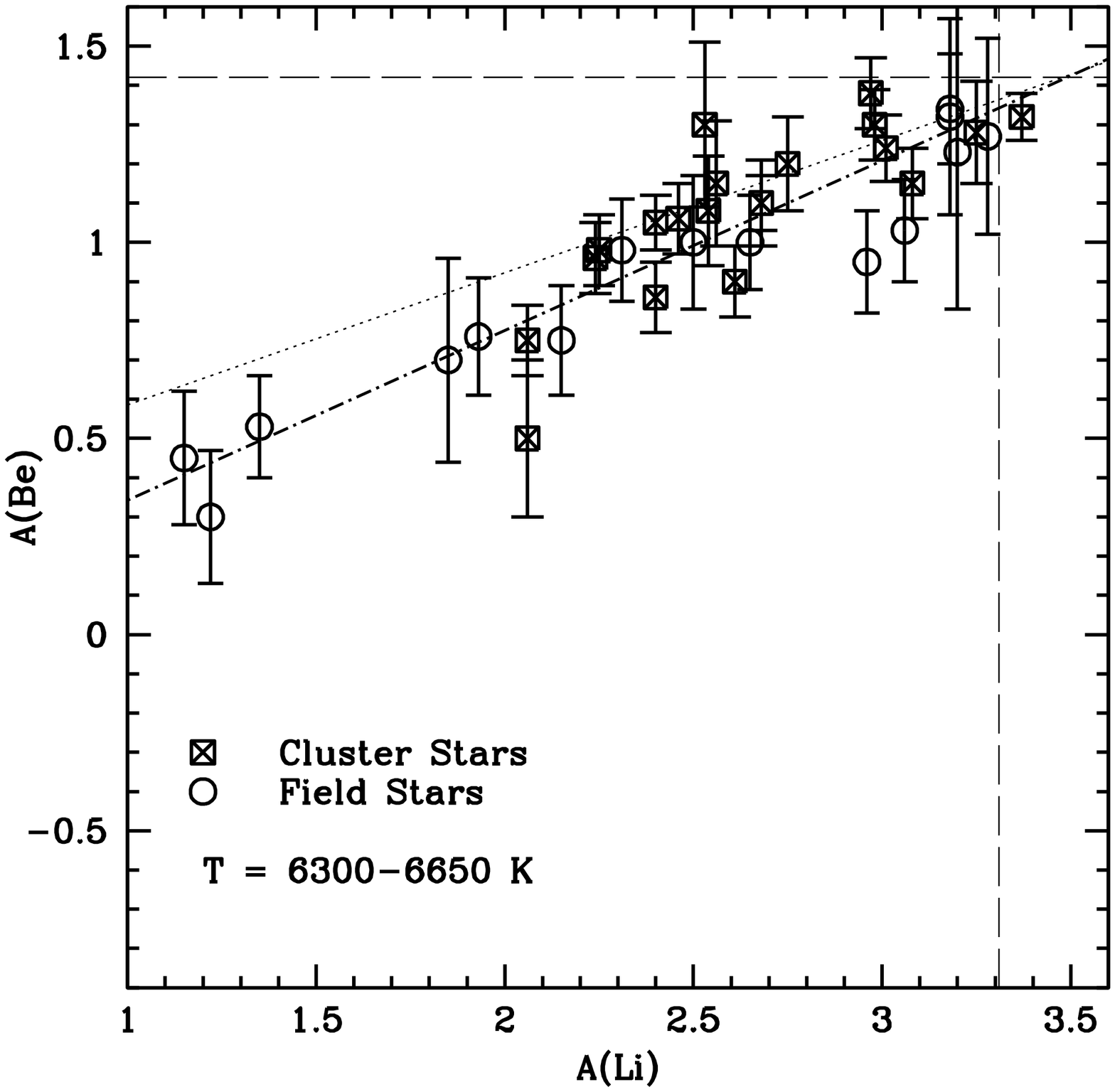}{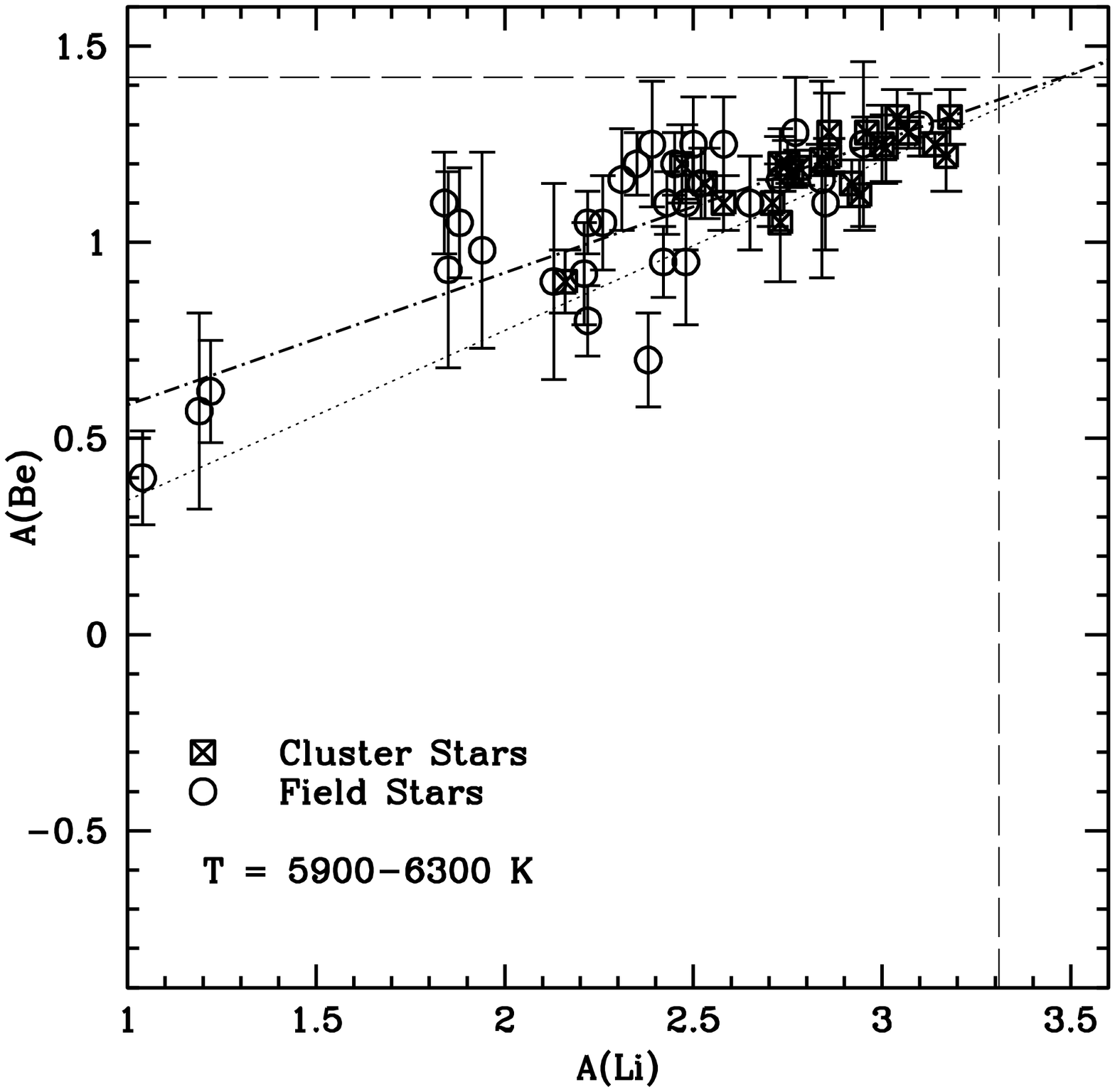}
\caption{Left: The Li and Be abundances in the hotter stars which corresponds
to the cool side of the Li-Be dip, 6300 - 6650 K.  The dot-dashed line
represents the best-fit least-squares slope of 0.43 $\pm$0.04.  The faint
dotted line is the slope for the stars in the right panel.  The horizontal and
vertical dashed lines correspond to the meteoritic Be and Li abundances
respectively.  Right: The Li and Be abundances in the cooler stars, 5900 -
6300 K.  The dot-dashed line represents the best-fit least-squares slope of
0.34 $\pm$0.03.  The faint dotted line is the slope for the stars in the left
panel.  ApJ, 613, 1202.}
\end{figure}

For the 35 stars with $T$ = 6300 - 6650 K the least squares fit gives the
relationship: A(Be) = 0.433 ($\pm$0.036) A(Li) $-$ 0.071 ($\pm$0.094)

For the 54 stars with $T$ = 5900 - 6300 K the least squares fit gives the
relationship: A(Be) = 0.337 ($\pm$0.031) A(Li) + 0.248 ($\pm$0.078)

The 88 stars in the total sample from $T$ = 5900 - 6650 K are shown in the
left panel of Figure 5.  The least squares solution for this line is:

A(Be) = 0.382($\pm$0.030) A(Li) + 0.105 ($\pm$0.078)   

\begin{figure}[!ht]
\plottwo{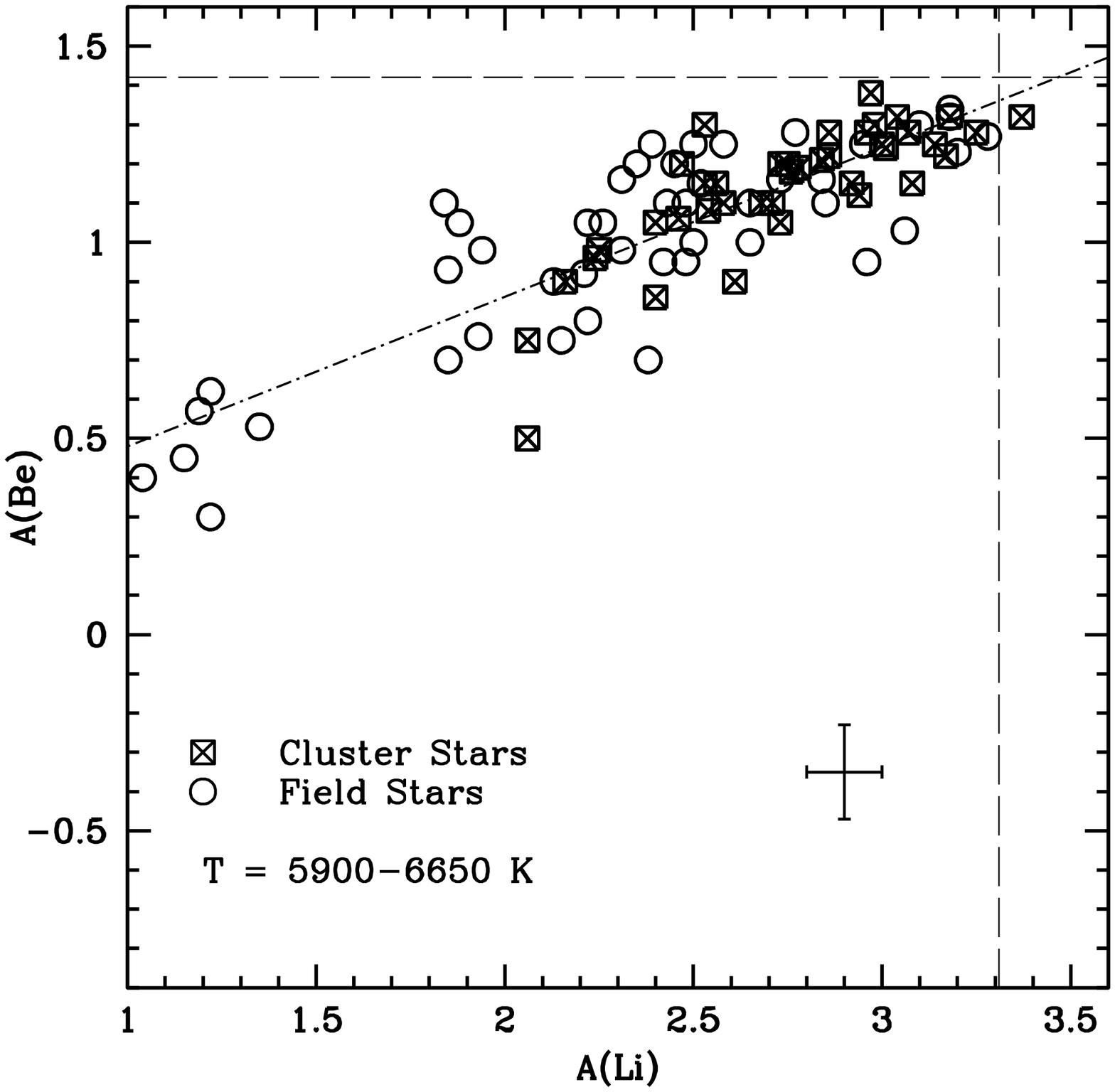}{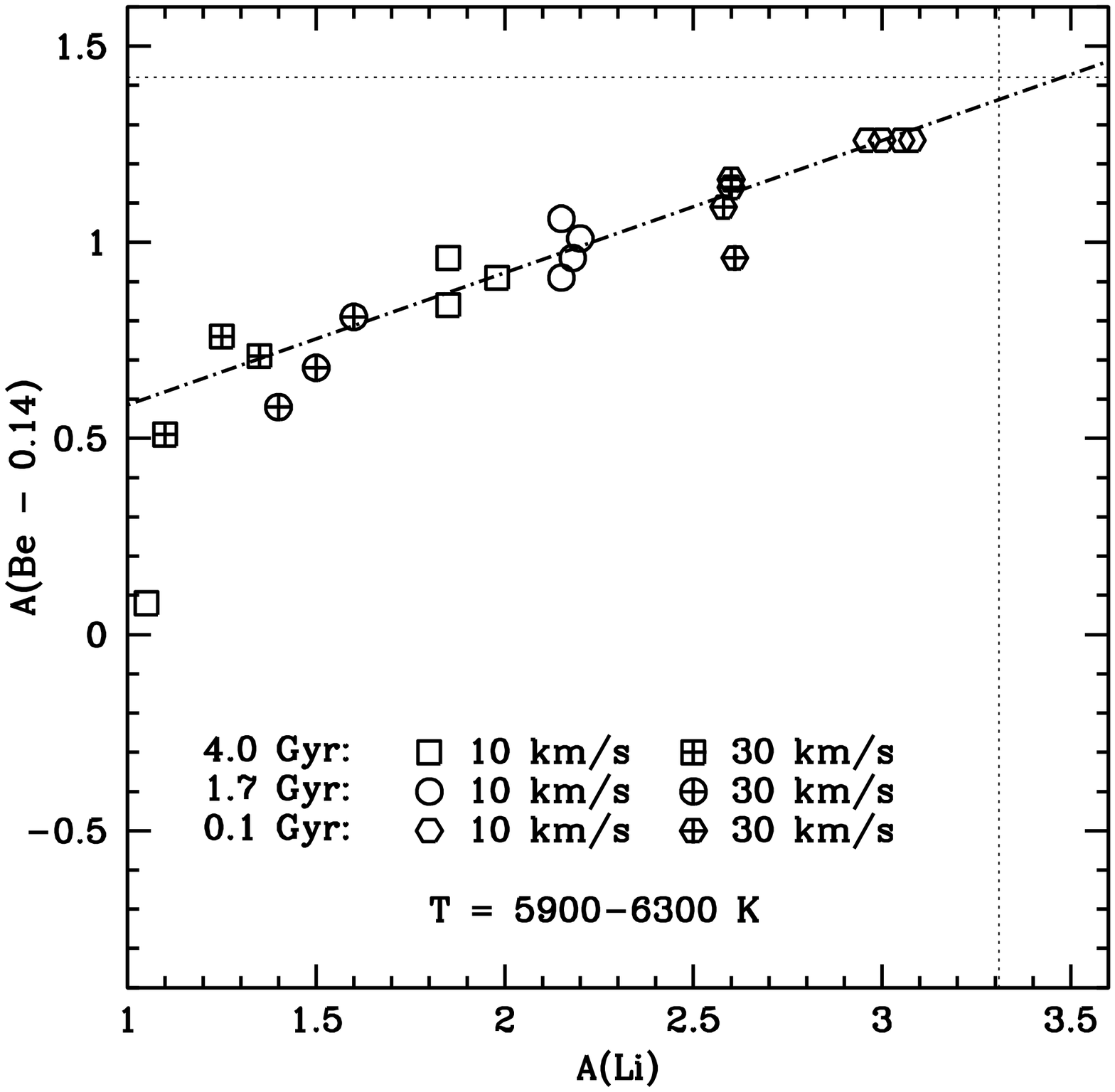}
\caption{Left: The Li and Be abundances in the full sample of 88 cluster and
field stars with $T$ = 5900 - 6650 K.  A typical error bar is in the lower
right.  The horizontal and vertical dashed lines represent the meteoritic Be
and Li abundances respectively.  Right: Models - see text.  ApJ, 613, 1202.}
\end{figure}

Calculations of Li and Be depletion due to rotationally-induced slow-mixing
have been made by Deliyannis \& Pinsonneault (1997) for dwarf stars of
selected ages (0.1, 1.7, and 4 Gyr), initial rotation (10 and 30 km s$^{-1}$)
and specific temperatures (see their Figure 2).  Examples of these depletions
at the three ages and the two initial velocities are shown in Figure 5, right
panel, for specific temperatures between 5900 and 6300 K.  The observed slope
is shown as the dot-dashed line from Figure 4, right panel.  For this
comparison the initial A(Be) is taken as 1.28, which corresponds to the mean
of the high values in Figure 4, left panel.  The observations and predictions
are in remarkably good agreement and very supportive of the conclusion that
the light element depletion is due to rotationally-induced mixing.  Deliyannis
\& Pinsonneault (1997) argue persuasively against mass loss and diffusion as
the source of the depletion.  

\subsection{Boron in Field Stars}

In order to study B in the Galactic disk, Boesgaard et al.~(2004c) used HST +
STIS to obtain spectra of the B I line at 2497 \AA\ in 20 stars which had no
depletion of Be.  The stars are from the upper envelope of Figure 3, right
panel, and cover a range in [Fe/H].  Since the Be abundance is undiminished in
these stars, the B abundances would be unaffected by depletion also.

\begin{figure}[!ht]
\plottwo{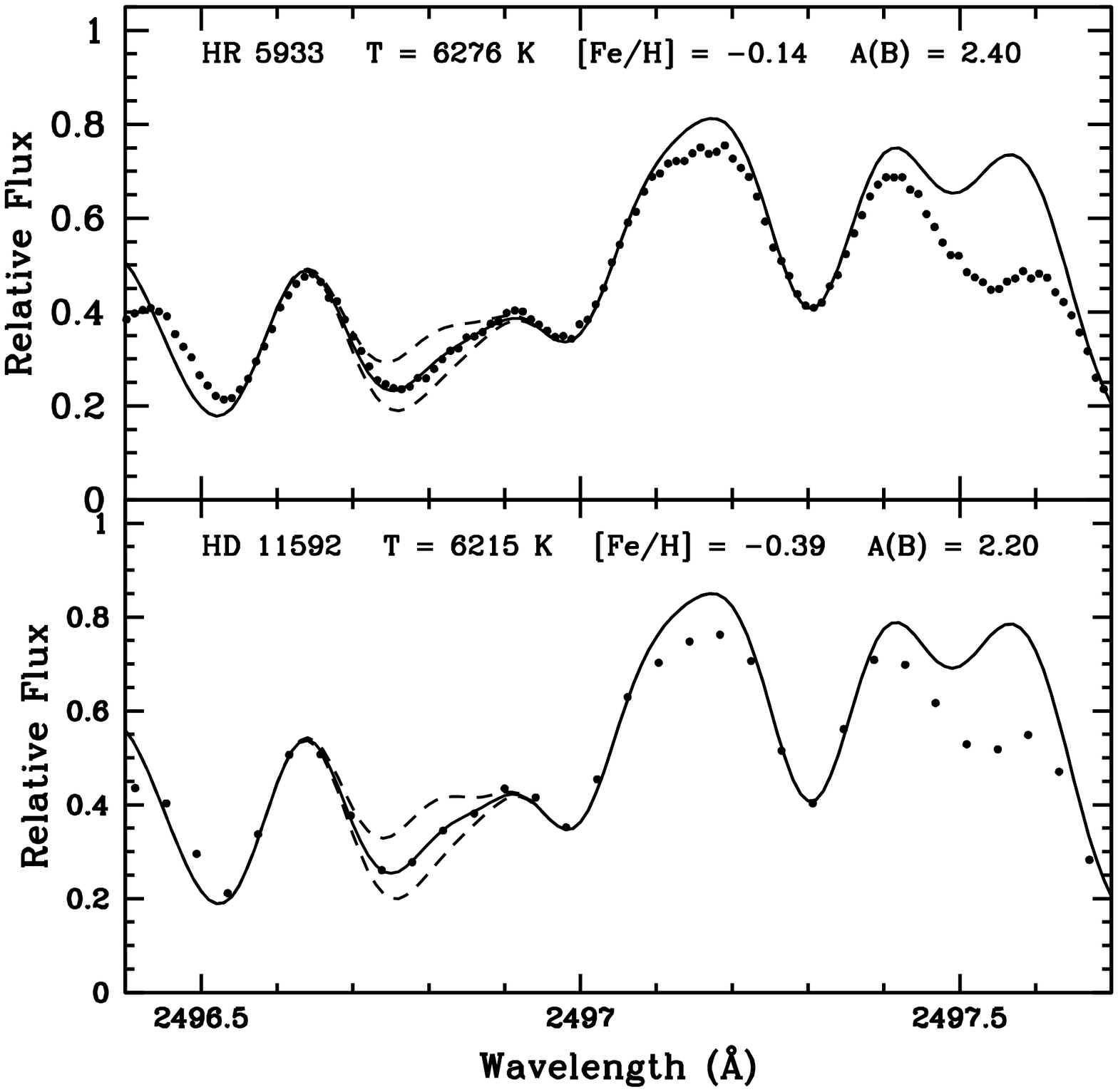}{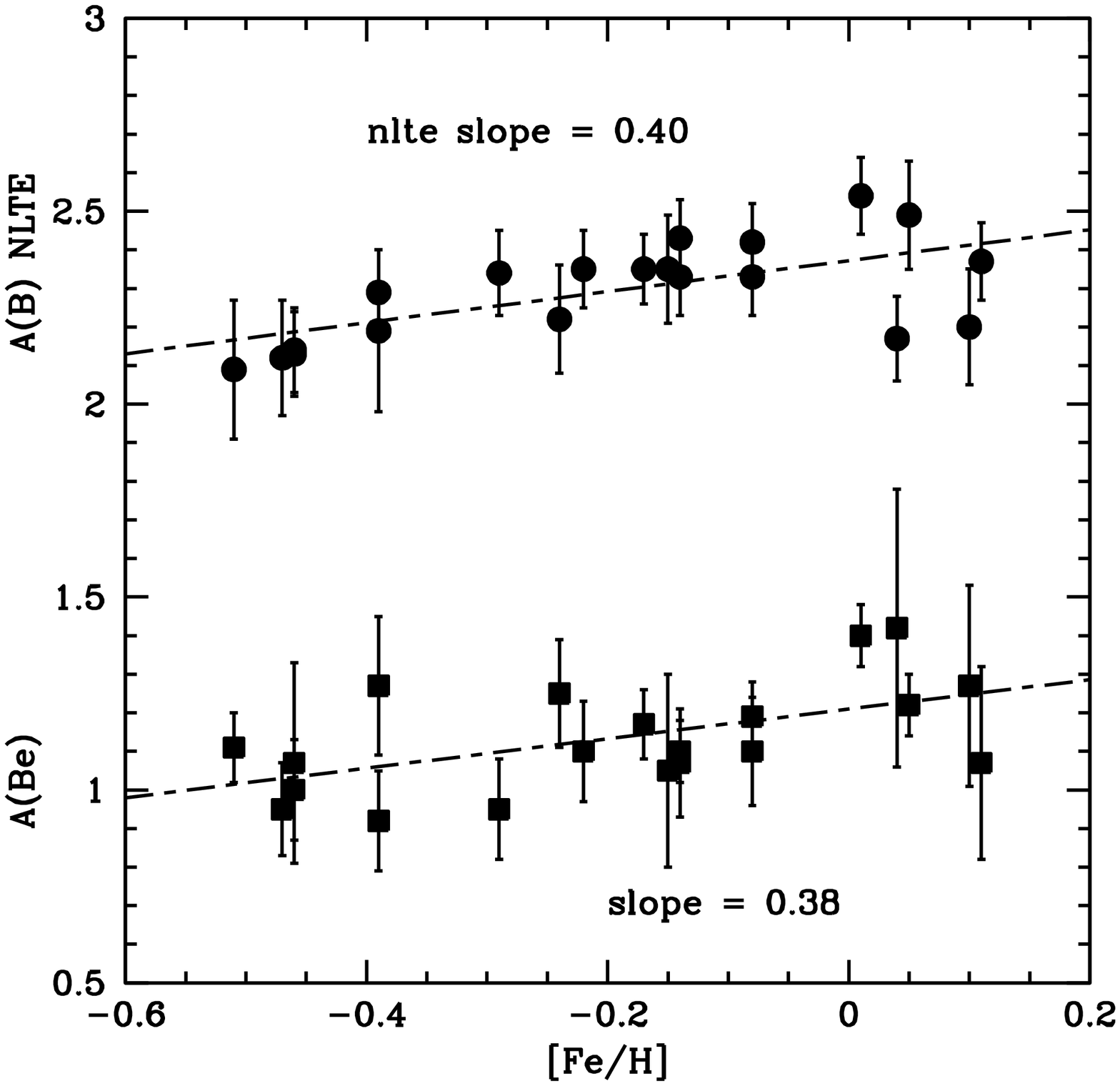}
\caption{Left: An example of the spectrum synthesis of the B I line.  The
upper star is a high resolution spectrum, $R$ = 114,000, while the lower star
is at medium resolution, $R$ = 30,000.  The dots are the observed points, the
solid line is the best fit, and the dashed lines are a factor of two in A(B)
above and below the best fit.  Right: The B and Be abundances in the sampl of
stars with undepleted Be.  There is a slope to the relationship with [Fe/H] of
about 0.40 for both Be and B.  ApJ, 606, 306.}
\end{figure}

Figure 6, left panel, shows and example of the spectrum synthesis for B I.
The results of the research to determine the initial values for B and Be in
Galactic disk stars are shown in Figure 6, right panel.  Both B and Be increase
slowly with Fe such that as Fe increases by a factor of 4, Be and B increase
by a factor of $\sim$1.7.

Those relationships are:

A(Be) = 0.382 ($\pm$0.135) [Fe/H] + 1.218 ($\pm$0.037)

A(B)$_{NLTE}$ = 0.402 ($\pm$0.117) [Fe/H] + 2.371 ($\pm$0.032)

The ratio of B/Be for the 20 stars was found to be 15, with no trend with
[Fe/H].  This ratio is consistent with the predictions of spallation reactions
as the source of Galactic Be and B.

\subsection{The Correlation of Be and B}

Another HST/STIS study of B focussed on stars with large Be depletions to
try to find stars that were depleted in B.  In this work Boesgaard, Deliyannis
\& Steinhauer (2005) looked at 13 F and G stars with large Be depletions and
at five Be-normal stars.  The result was the discovery of a correlation
between Be and B as shown in Figure 7.  The left panel showns the Be and B
abundances in our sample of 18 stars and has a slope of 0.20 $\pm$0.05.  Those
values have been corrected for the differences in the initial values of B and
Be that result from the relations seen in Figure 6, right panel.  Other Li-
and Be-normal stars have been added from the work described in the previous
section; those results are seen in Figure 7, right panel, where the slope is
0.18 $\pm$0.06.  Predictions for B depletion have not yet been made, but it
seems likely that rotational mixing will be a key ingredient in the small B
depletions. 

\begin{figure}[!ht]
\plottwo{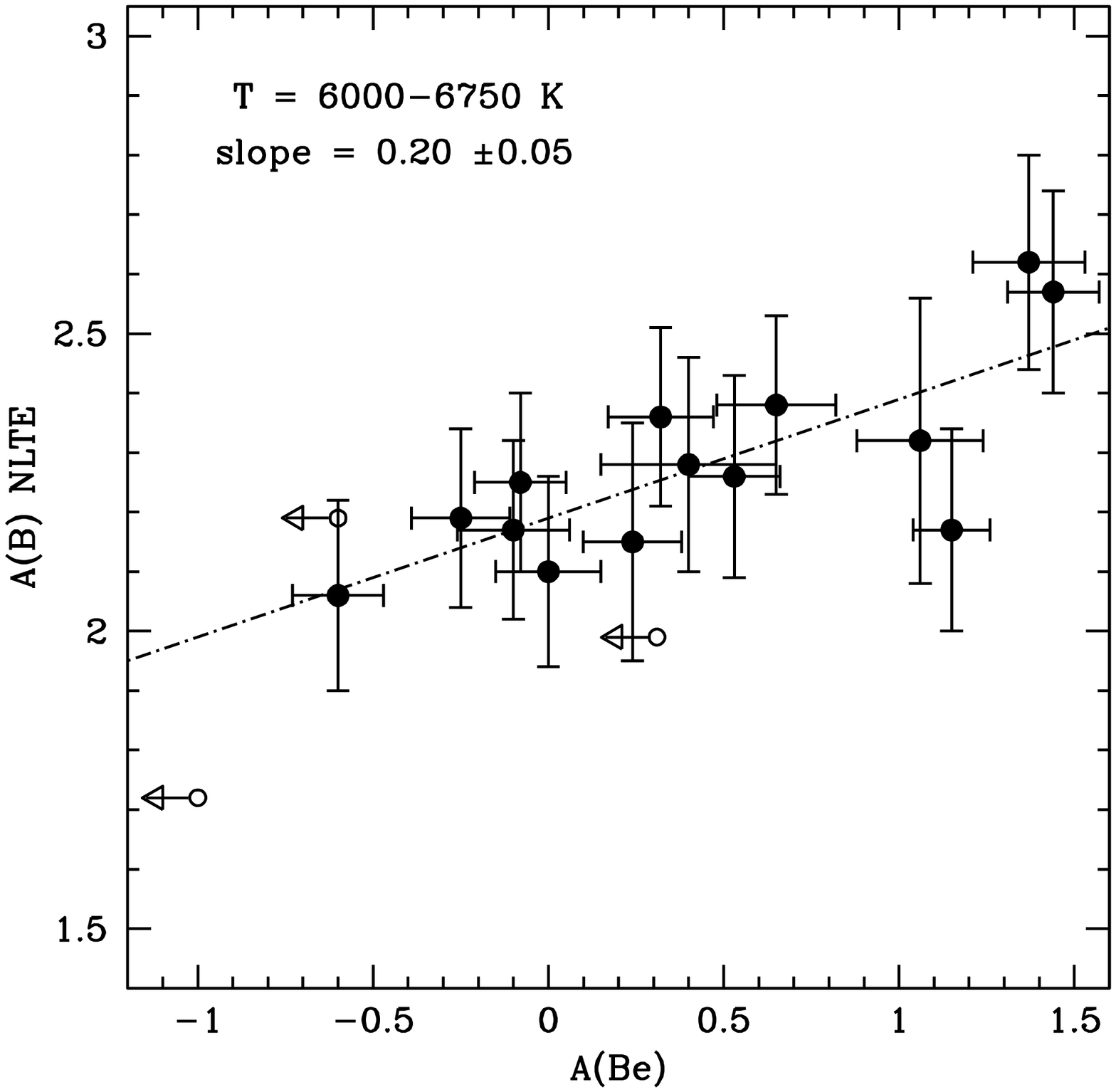}{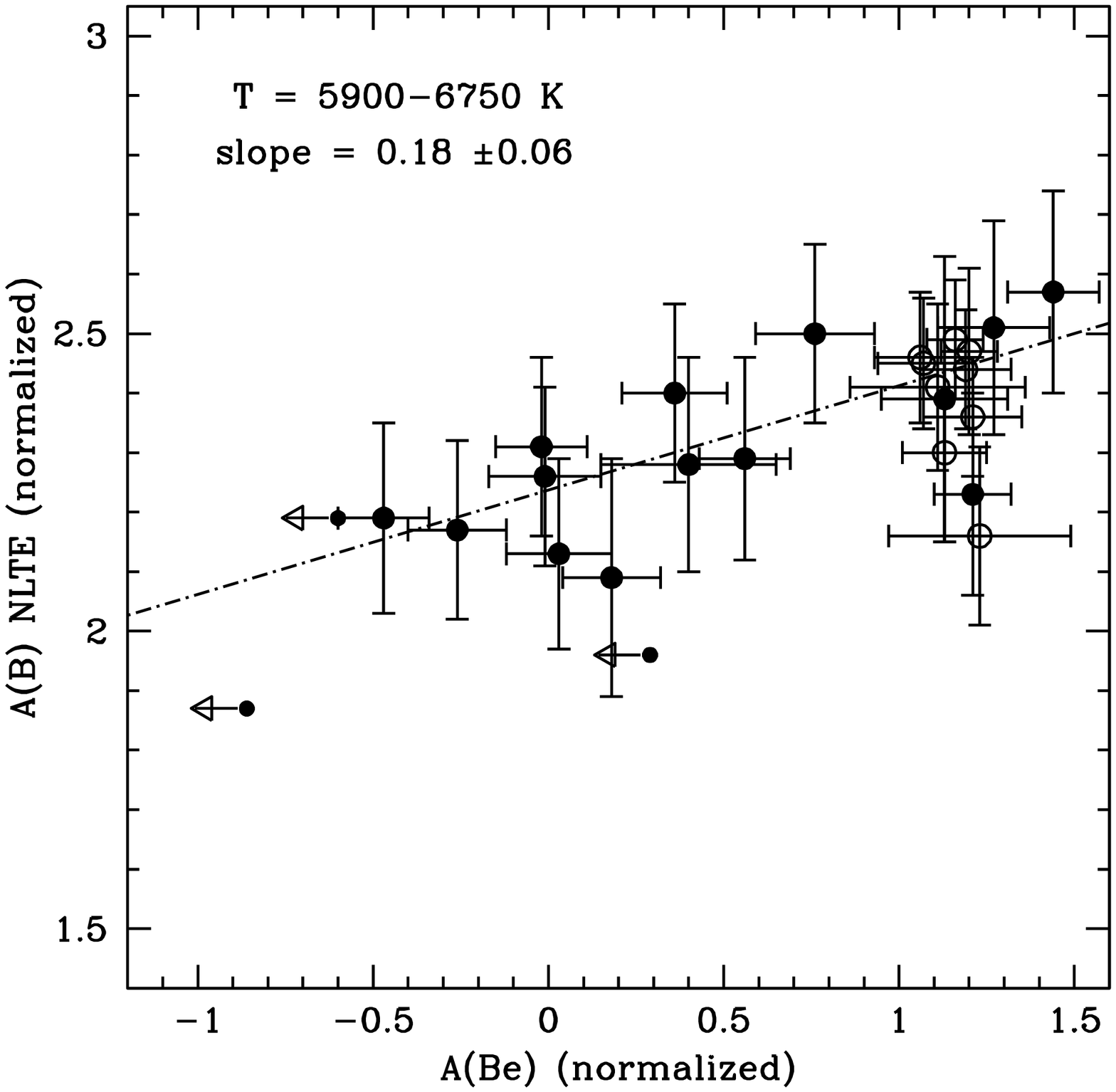}
\caption{Left: The correlation of B and Be.  Three stars with upper limits on
A(Be) are shown with left-directed arrows.  Right: The normalized values of
A(Be) and A(B) to correct for different initial values depending on [Fe/H], as
seen in Figure 6, right.  Additional Li-normal and Be-normal stars are
included.}
\end{figure}

\section{Li, Be, B Summary}

We have presented an amalgamation of several research projects on light
elements from spectra obtained primarily with the Keck telescope and HIRES and
HST with STIS.  In addition to the Li dip in several intermediate age
clusters, we have now discovered a Be dip in those clusters.  The Be dip is
not as deep as the Li dip, that is, Be is not as depleted as Li.  Neither Li
nor Be is measureably depleted in the younger open clusters, e.g.~Pleiades.
The depletion of these elements occurs during main-sequence evolution, after
an age of $\sim$100 Myr.  Although Li is depleted in cooler stars, and the
depletion increases with decreasing surface temperature, there is little or no
Be depletion in stars cooler than 5900 K.

Many stars in clusters and the field show depleted, but detectable, Li and
Be.  The abundances of Li and Be are found to be correlated in such stars with
Li being more depleted than Be.  For stars in the temperature range of 5900 -
6650 K the relationship is: 

A(Be) = 0.382 ($\pm$0.030) A(Li) + 0.105 ($\pm$0.078).

\noindent
When Li has decreased by a factor of 10, Be has decreased by only a factor of
2.4.  This relationship is very well matched by the predictions of models with 
rotationally-induced mixing.

Through use of B abundances from HST data we have found that there are stars
with depleted, but detected, B.  And there is a correlation between the B and
Be abundances.  That relationship is:

A(B) = 0.175 ($\pm$0.058) A(Be) + 2.237 ($\pm$0.055).

\noindent
This slope is shallower than the Li-Be slope because the B preservation region
is larger and B atoms have to be mixed to deeper stellar layers to be 
depleted.  When Be is down by a factor of 10, B is only reduced 1.5 times.

Li, Be, B all seem to increase at about the same rate with [Fe/H] in Galactic
disk stars.  The slope for A(B) with [Fe/H] is 0.40 and for A(Be) with [Fe/H]
is 0.38.  Large depletions have been found for all three elements in our
samples of F and G Population I stars.

\section{Carbon and Oxygen in Open Clusters}

After H and He, the most abundant elements are C, N, O, and Ne.  Determining
those abundances, however, is a difficult proposition.  This is a report on an
ongoing project to determine abundances in open clusters, and focuses on the
determination of Fe, C, and O abundances in three clusters: the Pleiades at an
age of $\sim$70 Myr, the Hyades at $\sim$700 Myr and M 67 at 5 Gyr.  The
spectra for all were obtained with Keck I + HIRES at high S/N and high
spectral resolution of $\sim$48,000.  The spectral range covered 5700 - 8120
\AA.  Abundances were determined from some 40 lines of Fe I, 6 of C I
and the 3 of the O I triplet.

\subsection{Oxygen Abundances}

Although abundances found from the O I triplet at 7771, 7773, and 7774 \AA\
are influenced by the effects of NLTE, the corrections to be applied are
complicated to determine and various researchers have found different results,
ranging from a few hundredths to a few tenths of a dex.  (In particular, there
is uncertainty in the size of the collision rates and cross sections for
collisions with neutral hydrogen, e.g.~Kiselman 2001.)  It is possible to take
an empirical approach on this issue.  First, we can determine O abundances in
a number of stars of different temperatures on the main sequence in the same
cluster.  All unevolved stars in a given open cluster can be expected to have
the same O abundance.  We have done this for 30 stars in the Hyades cluster in
a temperature range of 5700 - 7100 K; the spectra are from the Palomar 5-m
coud\'e and Keck + HIRES.  The O abundances (on the scale where log (H) =
12.00) from the O I triplet as a function of effective temperature are shown
in Figure 8 (left).  While the hotter stars show large O abundances, the
values found from Hyades stars cooler than 6200 K are consistent with each
other.  This suggests that the NLTE effects are small (at least consistent)
in the cool stars.

\begin{figure}[!ht]
\plottwo{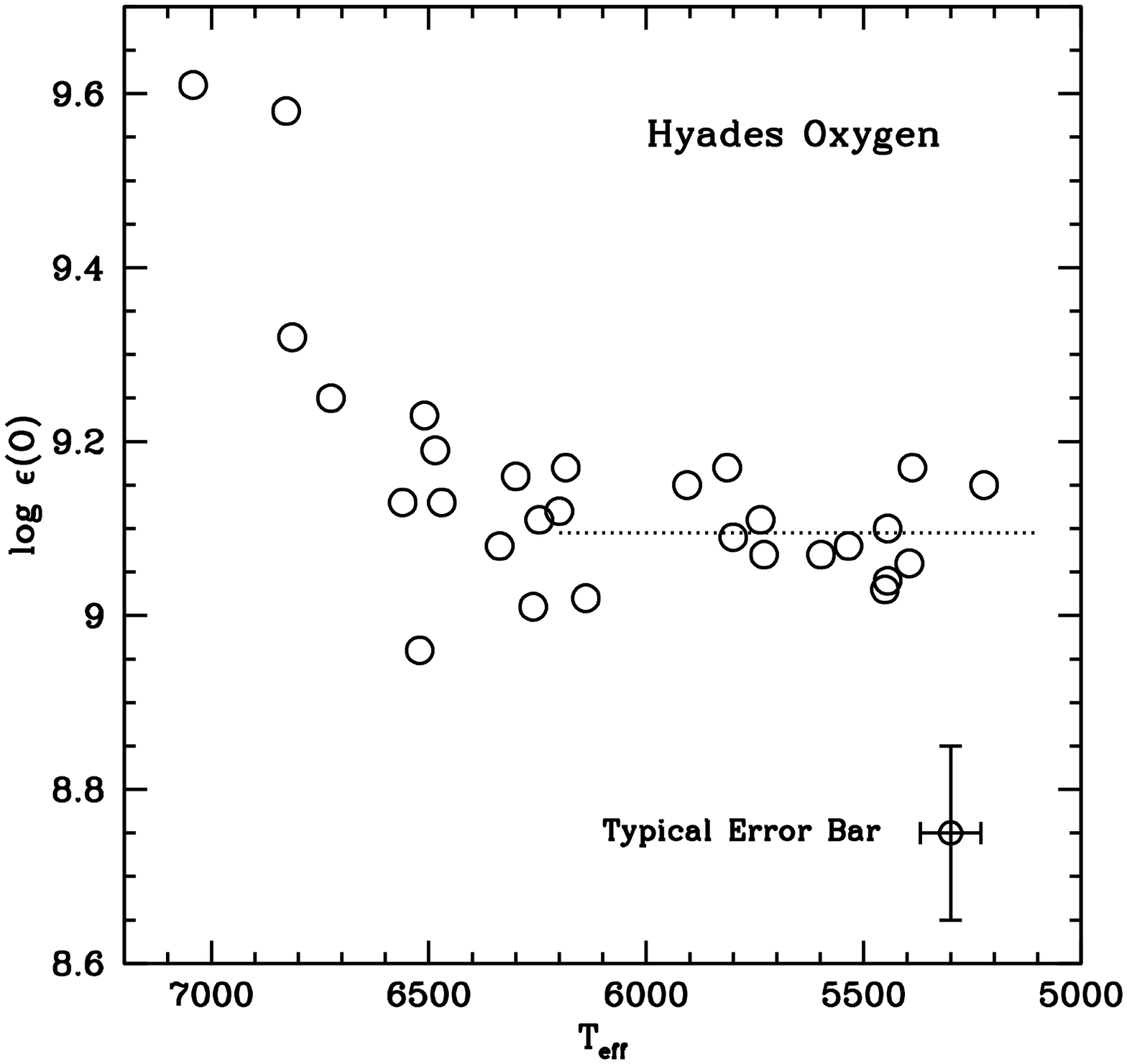}{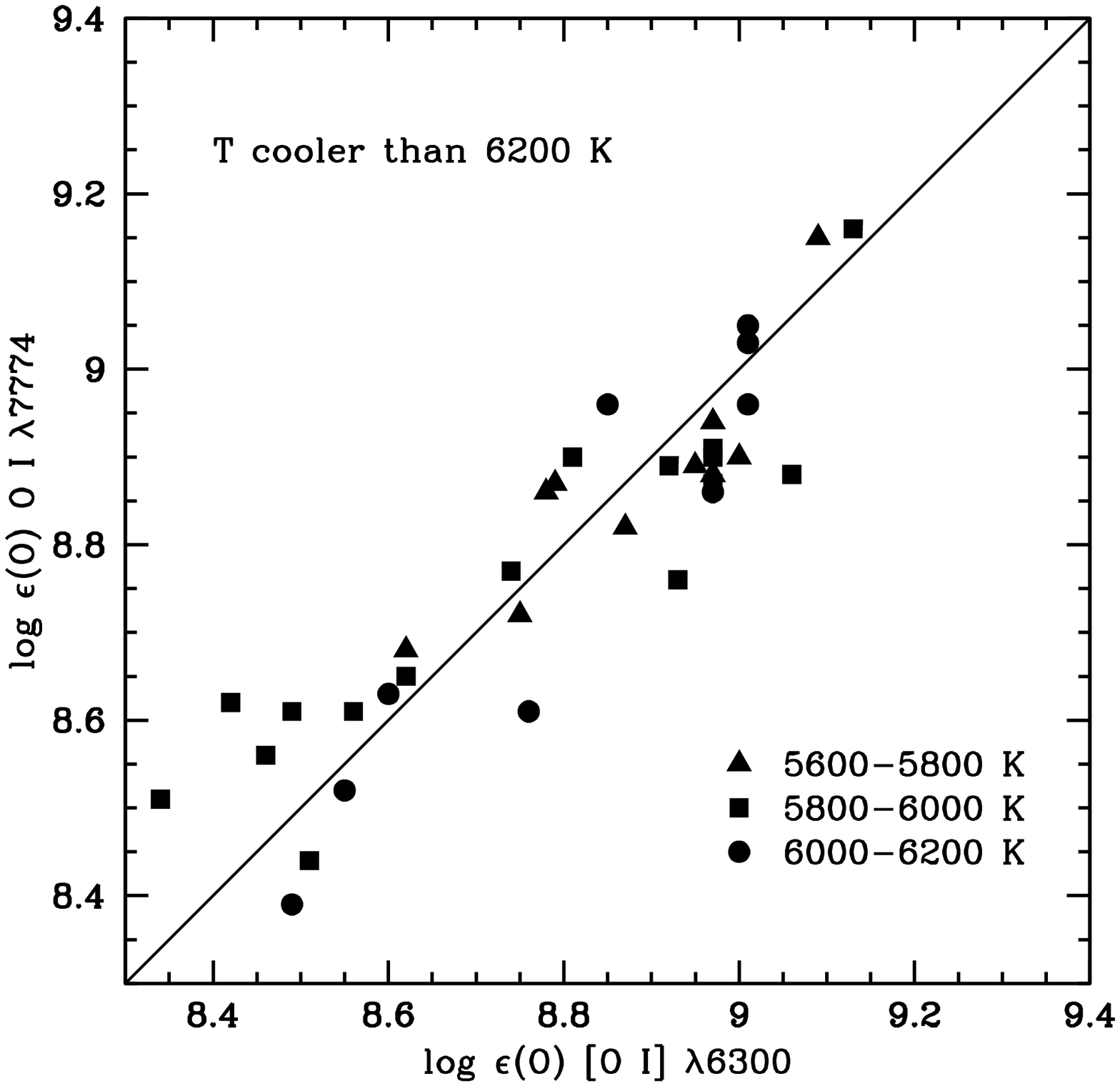}
\caption{Left: Abundances of O from the O I triplet in Hyades main sequence
stars, showing the effect of stellar temperature.  All stars in the Hyades
should have the same O abundance, but the hotter stars give abundances that
are higher than the cooler stars, presumably due to effect of NLTE.  The
horizontal dotted line corresponds to a uniform value of log N(O)/N(H) + 12.00
= 9.10 for the 15 stars with $T_{\rm eff}$ $<$ 6200 K.  Right: Abundances of O
from observations of the O I triplet and the [O I] forbidden line in the same
stars, all having $T_{\rm eff}$ less than 6200 K.  The diagonal line shows the
equivalent abundance from the two techniques.  The stars are subdivided into
three temperature groups.  No differences are found in the three groups.  The
abundance errors are typically $\pm$0.10.}
\end{figure}

Another empirical approach is to examine O abundances found from the O I
triplet and the forbidden [O I] line at 6300 \AA\ when these features can be
measured in the same stars.  This can be done with the data set in King \&
Boesgaard (1995).  The result of this is shown in Figure 8 (right) for stars
cooler than 6200 K.  The diagonal line corresponds to the same abundance from
the two different features of O.  The two methods give the same result (within
the typical error of $\pm$0.10 dex) with no apparent difference in the three
temperature groupings shown in the figure.

As a result of these two empirical demonstrations, we have determined O
abundances from the O I triplet from observations of main sequence stars
cooler than 6200 K.  Note that Schuler et al.~(2004) find that O abundances
from the O I triplet begin to increase again for stellar temperatures below
5400 -- 5500 K in their study of O abundances in the Pleiades and M 34.  The
stars in our samples in the Pleiades and M 67 are all hotter than
5500 K.

\subsection{Hyades}

We have found abundances in some 18 Hyades late-F and early-G dwarfs from Keck
HIRES spectra with S/N from 610 to 730.  Figure 9 (left) shows the positions
of these stars in the Hyades HR diagram.  The abundances of Fe were 

\begin{figure}[!ht]
\plottwo{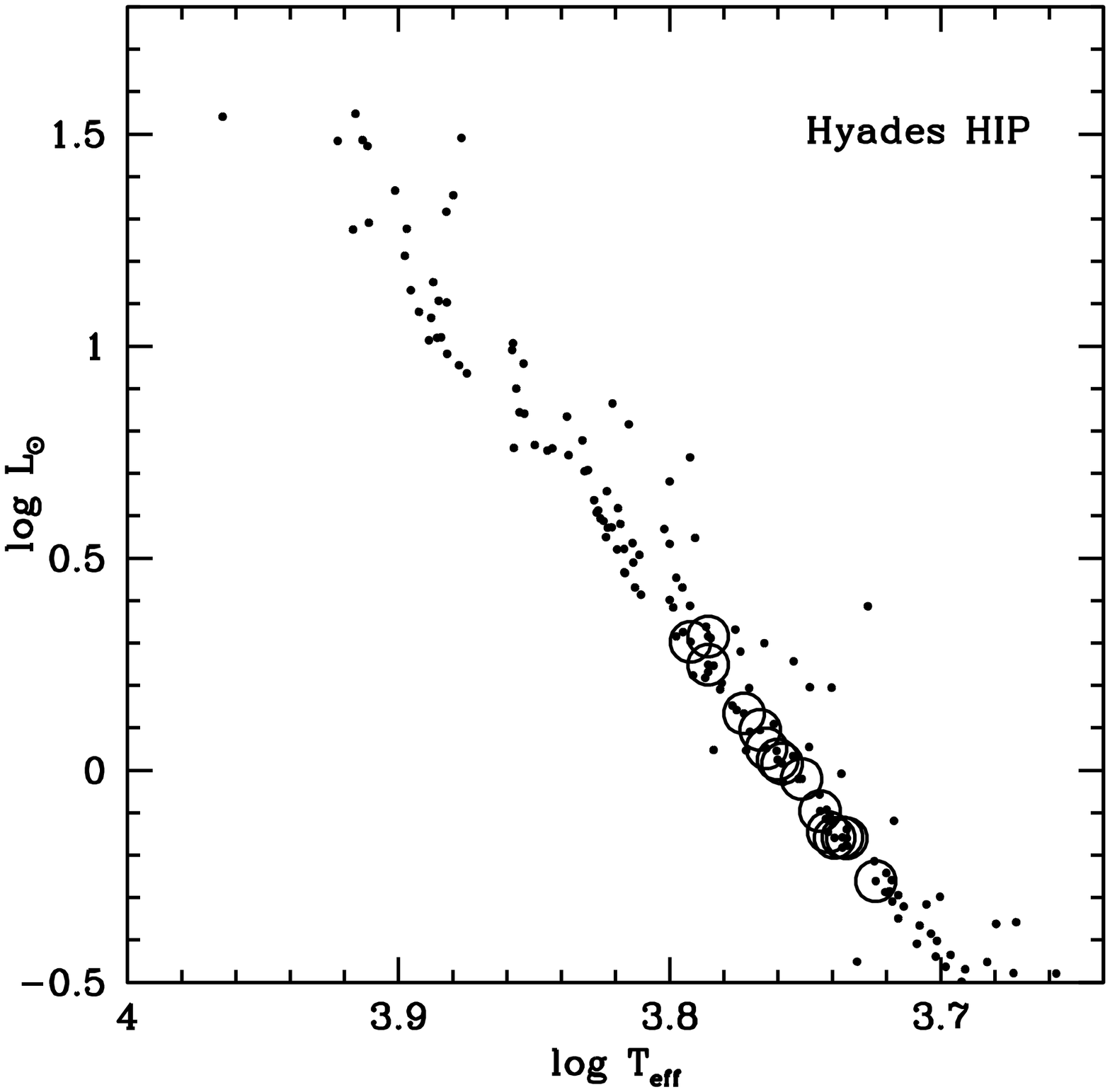}{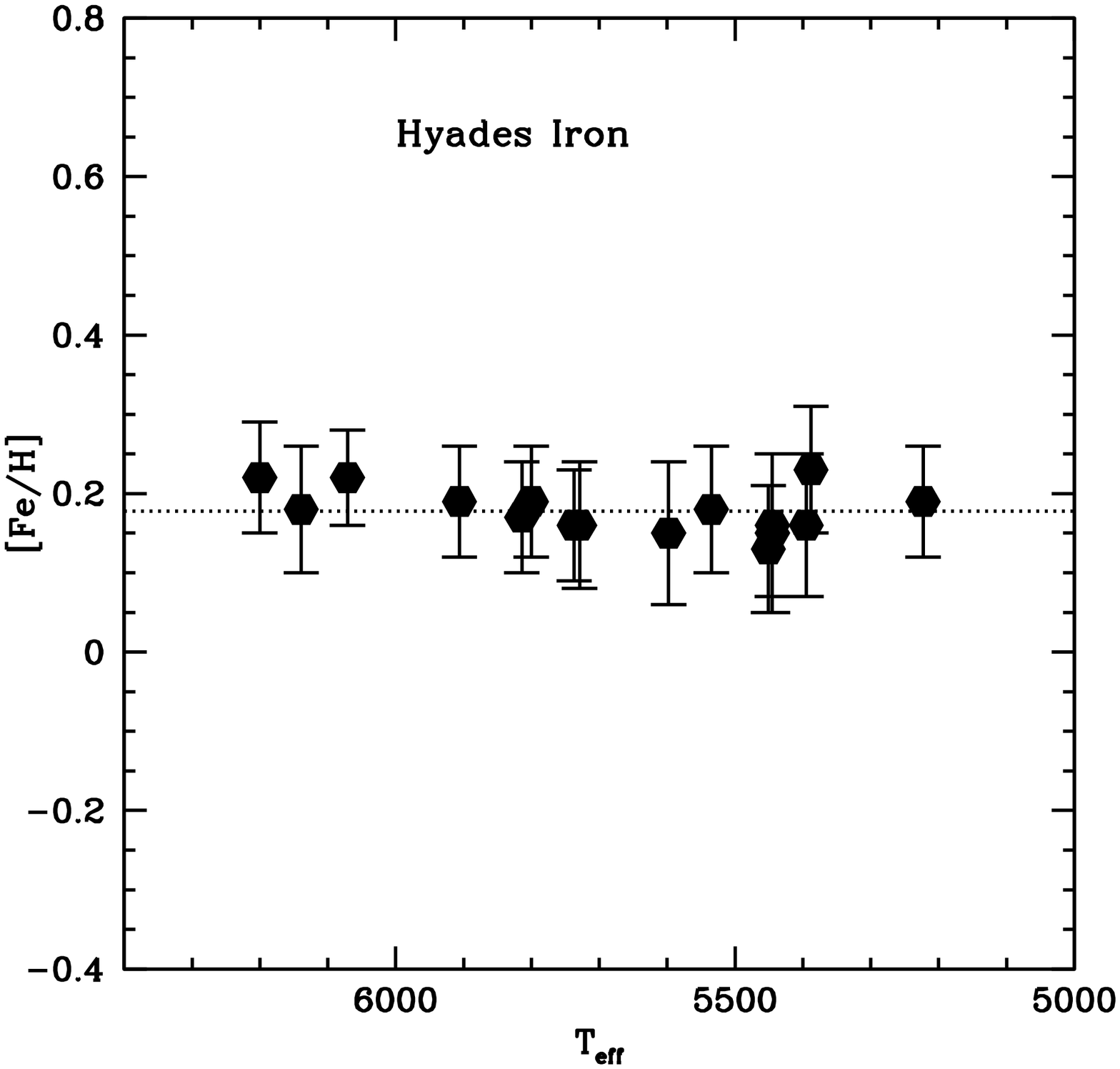}
\caption{Left: HR diagram for the Hyades from Hipparcos data.  The stars we
observed are shown as large open circles.  Right: Abundances of Fe in the
Hyades stars.  The horizontal dotted line shows the cluster mean [Fe/H] =
+0.18 $\pm$0.01.}
\end{figure}

\noindent
found from $\sim$40 Fe I lines for each star; those results are shown in the
right panel of Figure 9.  The mean cluster abundance is [Fe/H] = +0.18
$\pm$0.01, in good agreement with the recent value of +0.16 $\pm$0.02 (Yong et
al.~2004).

\begin{figure}[!ht]
\plottwo{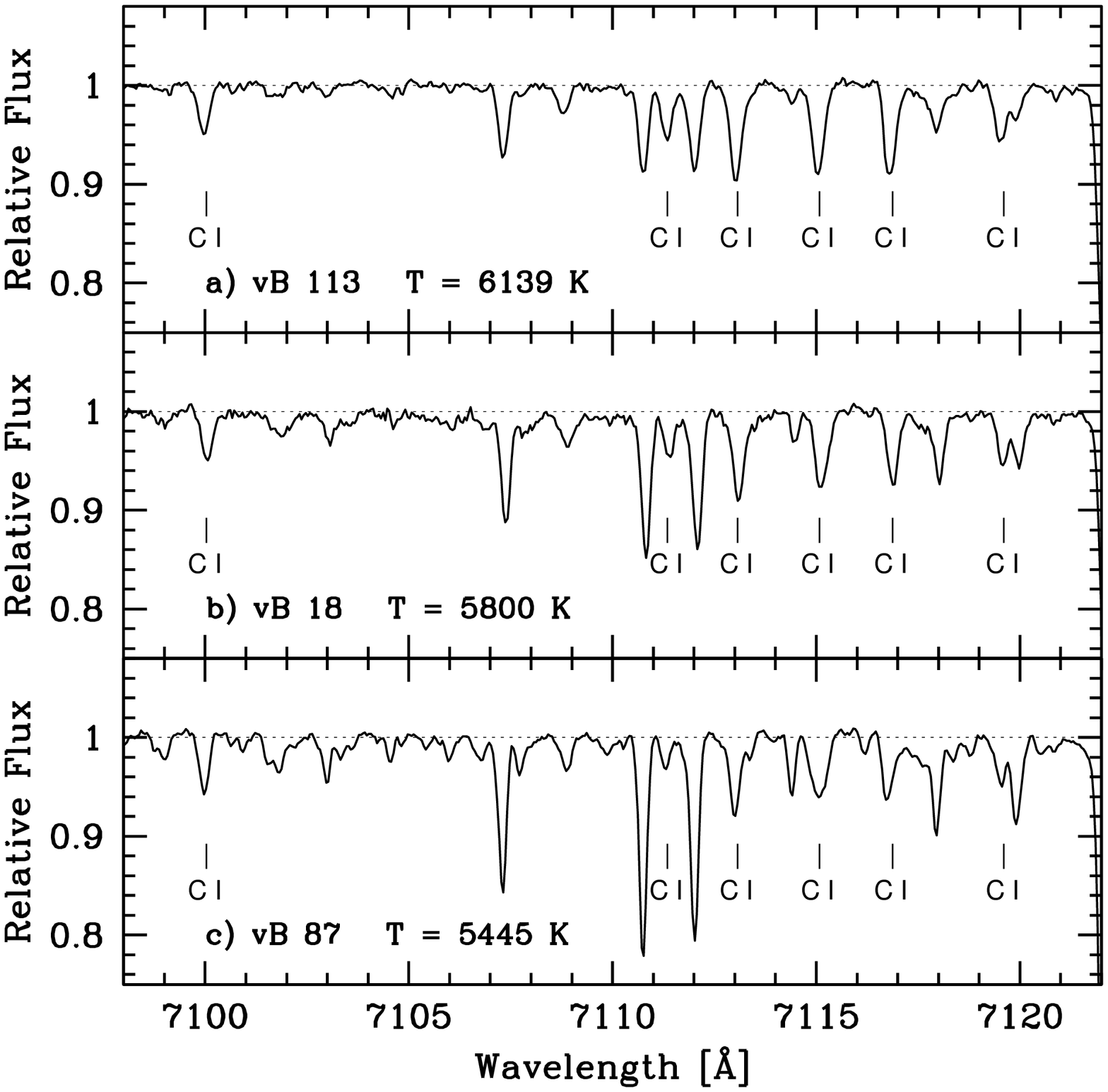}{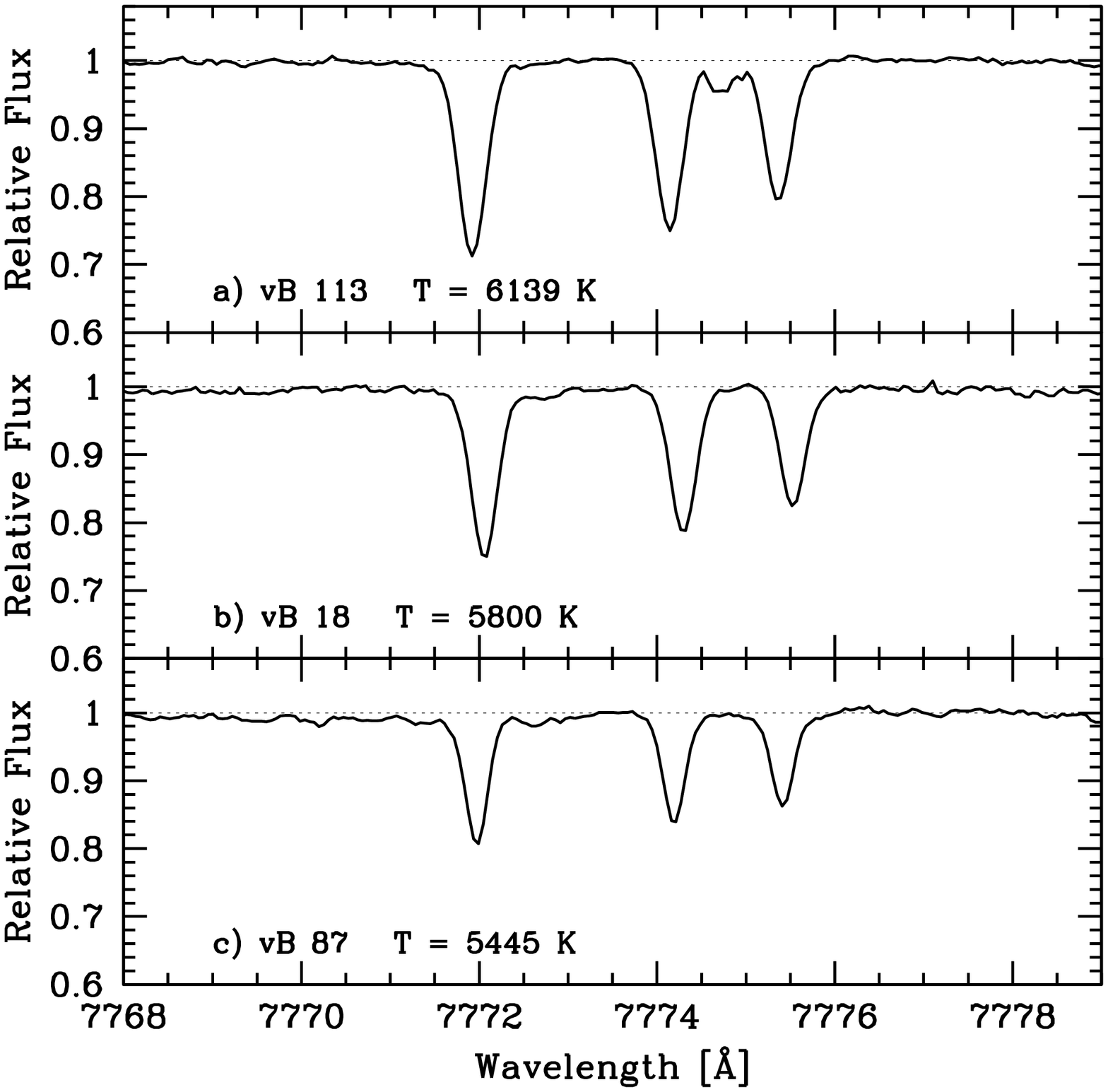}
\caption{Examples of the spectra of the Hyades in the C I and the O I regions.
The top star is one of the hottest in our sample, the middle one is near the
solar temperature and the lower one is one of the coolest in the sample.}
\end{figure}

We have found O and C abundances from the high excitation lines of these
elements in the near IR spectrum.  The lines used are shown in each of three
Hyades stars in Figure 10.  The abundances are shown as a function of
temperature in Figure 11.  The stars studied are all cooler than 6200 K.
For the mean cluster [C/H] we find +0.16 $\pm$0.02 and a mean [C/Fe] =
$-$0.02.  For [O/H] we find the mean cluster abundance to be +0.17 $\pm$0.01
and a mean [O/Fe] = $-$0.01.

\begin{figure}[!ht]
\plottwo{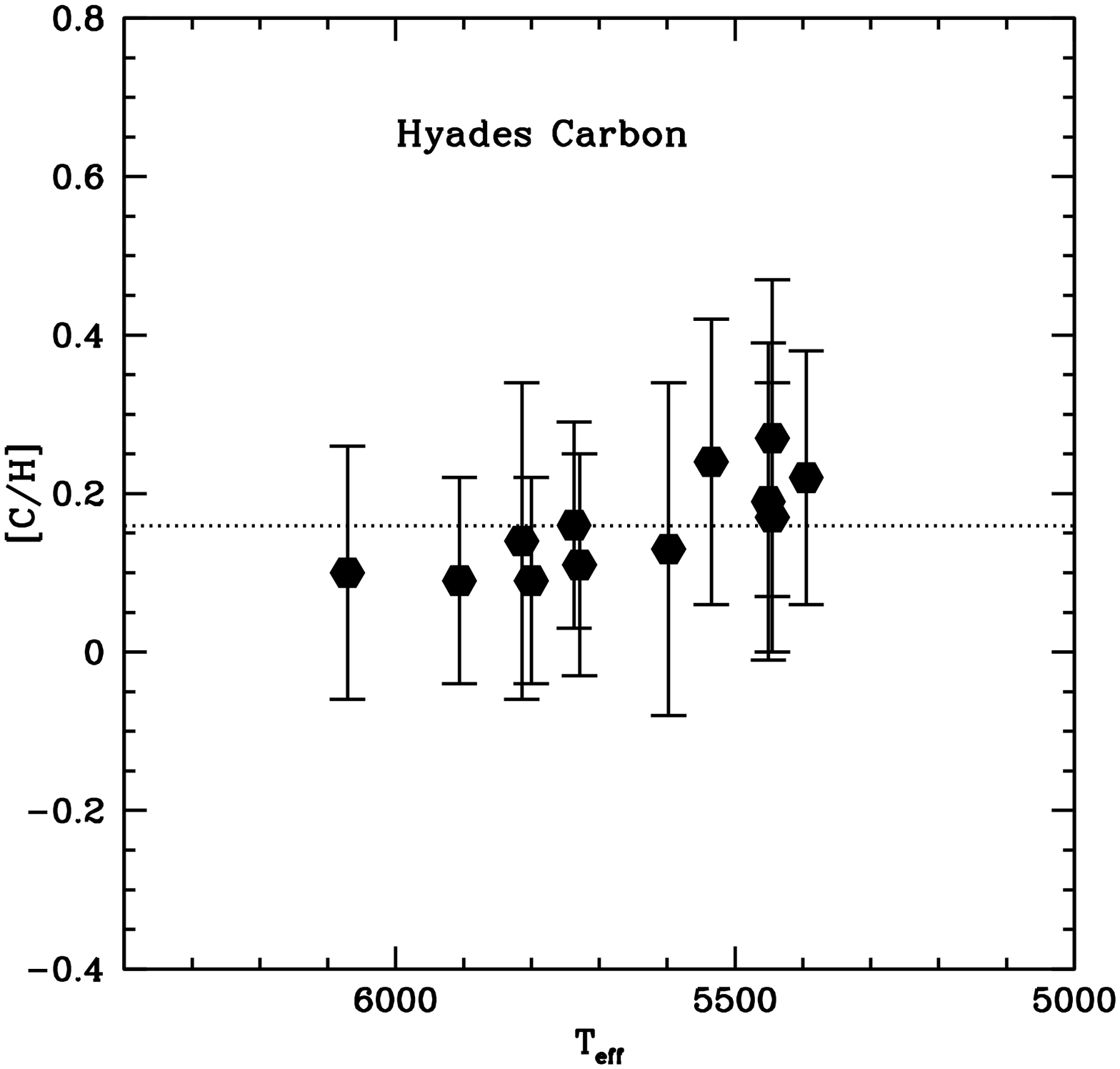}{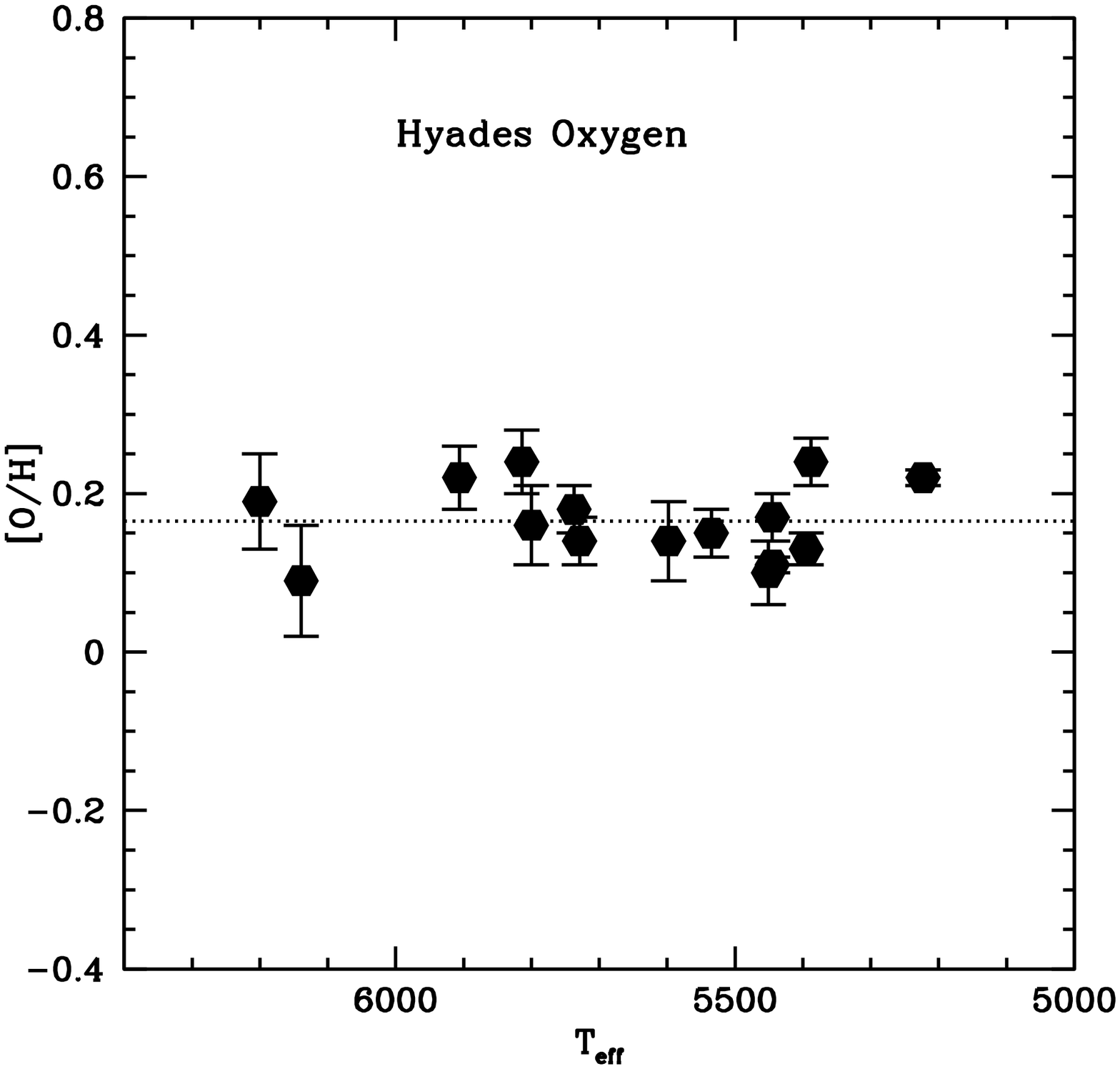}
\caption{Abundance results for the Hyades for C (left panel) and O (right
panel).  The horizontal dotted lines show the cluster means: [C/H] = +0.16
$\pm$0.02 and [O/H] = +0.17 $\pm$0.02.}
\end{figure}

\subsection{Pleiades}

We have obtained Keck/HIRES spectra of 20 late-F and early-G dwarfs with S/N
ratios of 95 -- 160.  The color-magnitude diagram for the Pleiades is shown in
the left panel of Figure 12.  The right panel of Figure 12 shows the results
for [Fe/H] in these Pleiades stars from $\sim$40 lines.  

\begin{figure}[!ht]
\plottwo{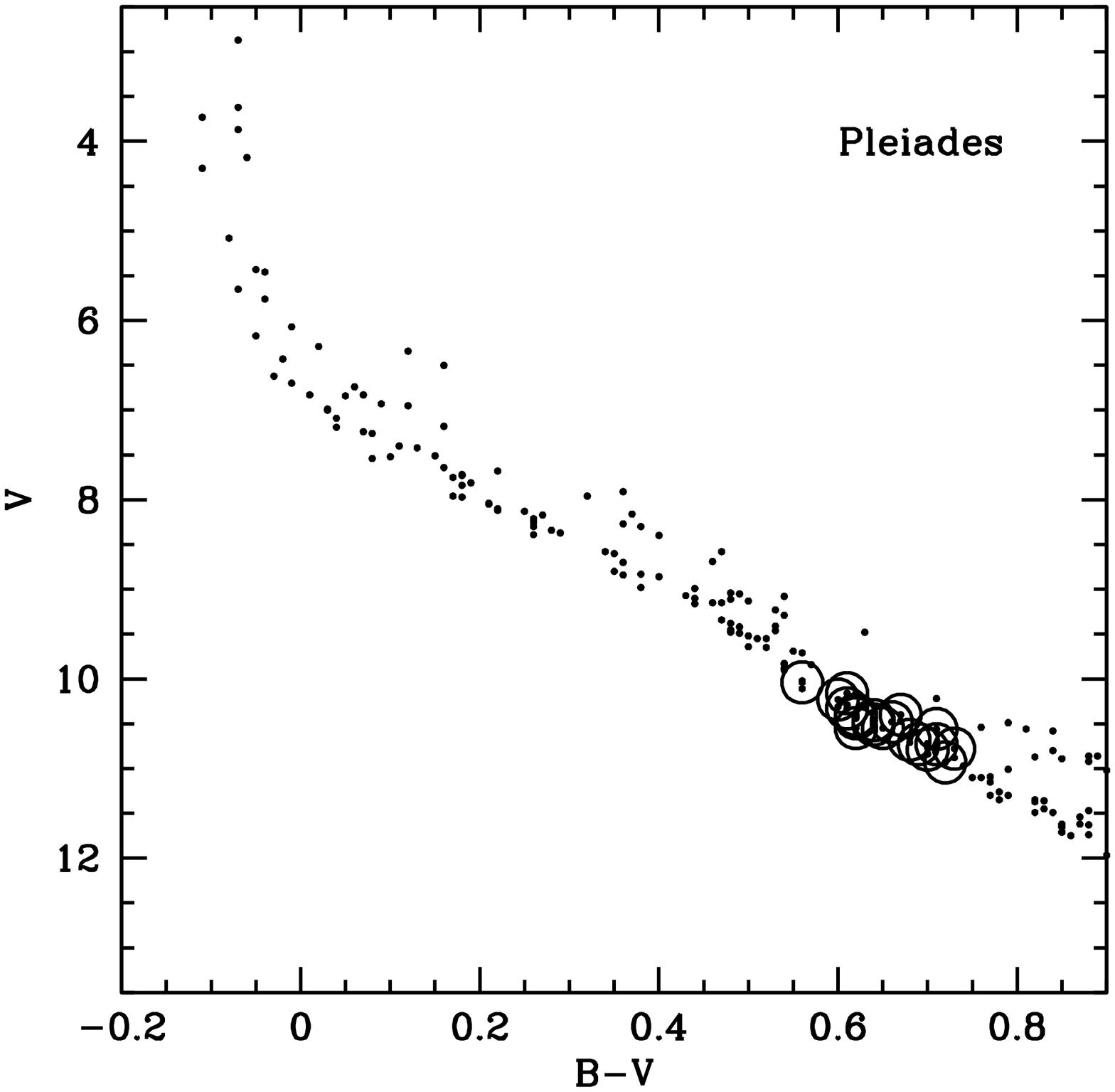}{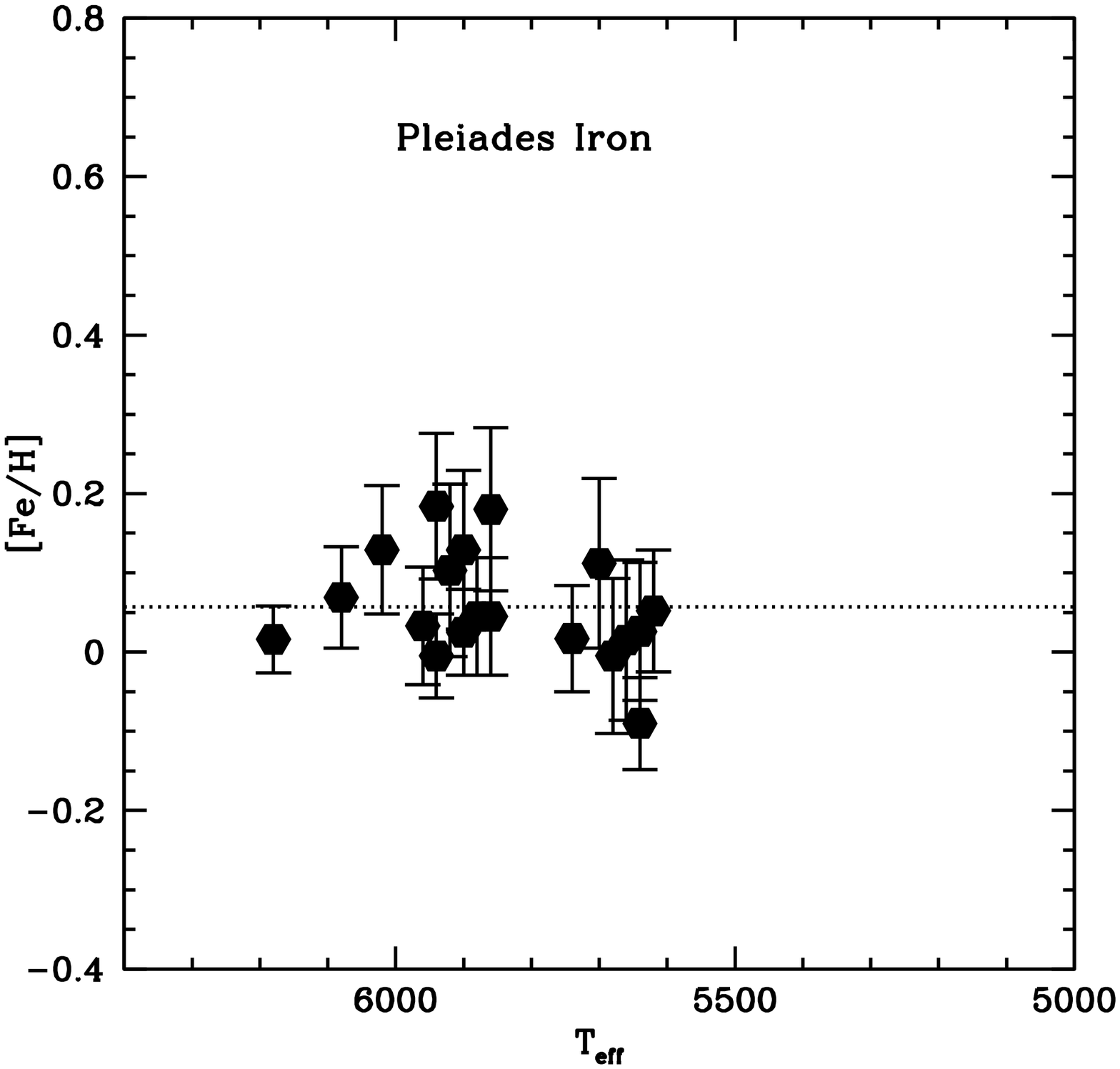}
\caption{Left: Color-magnitude diagram for the Pleiades with the large crosses
indicating the stars observed.  Right: Results for the Fe abundances in the
Pleiades stars.  The cluster mean [Fe/H] = +0.06 $\pm$0.02 (dotted line).}
\end{figure}

The mean value for the cluster is [Fe/H] = +0.06 $\pm$0.02, which is slightly
higher than previous values of +0.02 $\pm$0.06 (Boesgaard 1989), $-$0.03
$\pm$0.02 (Boesgaard \& Friel 1990), 0.00 $\pm$0.05 (Boesgaard, Budge \&
Ramsay 1988).  The abundance results for C and O are shown graphically in
Figure 13.  For [O/H] we find +0.11 $\pm$0.02 and thus [O/Fe] = +0.05.  Our
[O/H] is in good agreement with that from [O I] $\lambda$6300 of +0.14 from
Schulaer et al.~(2004).  Carbon is low in the Pleiades with [C/H] = $-$0.04
$\pm$0.02 and [C/Fe] = $-$0.10.

\begin{figure}[!ht]
\plottwo{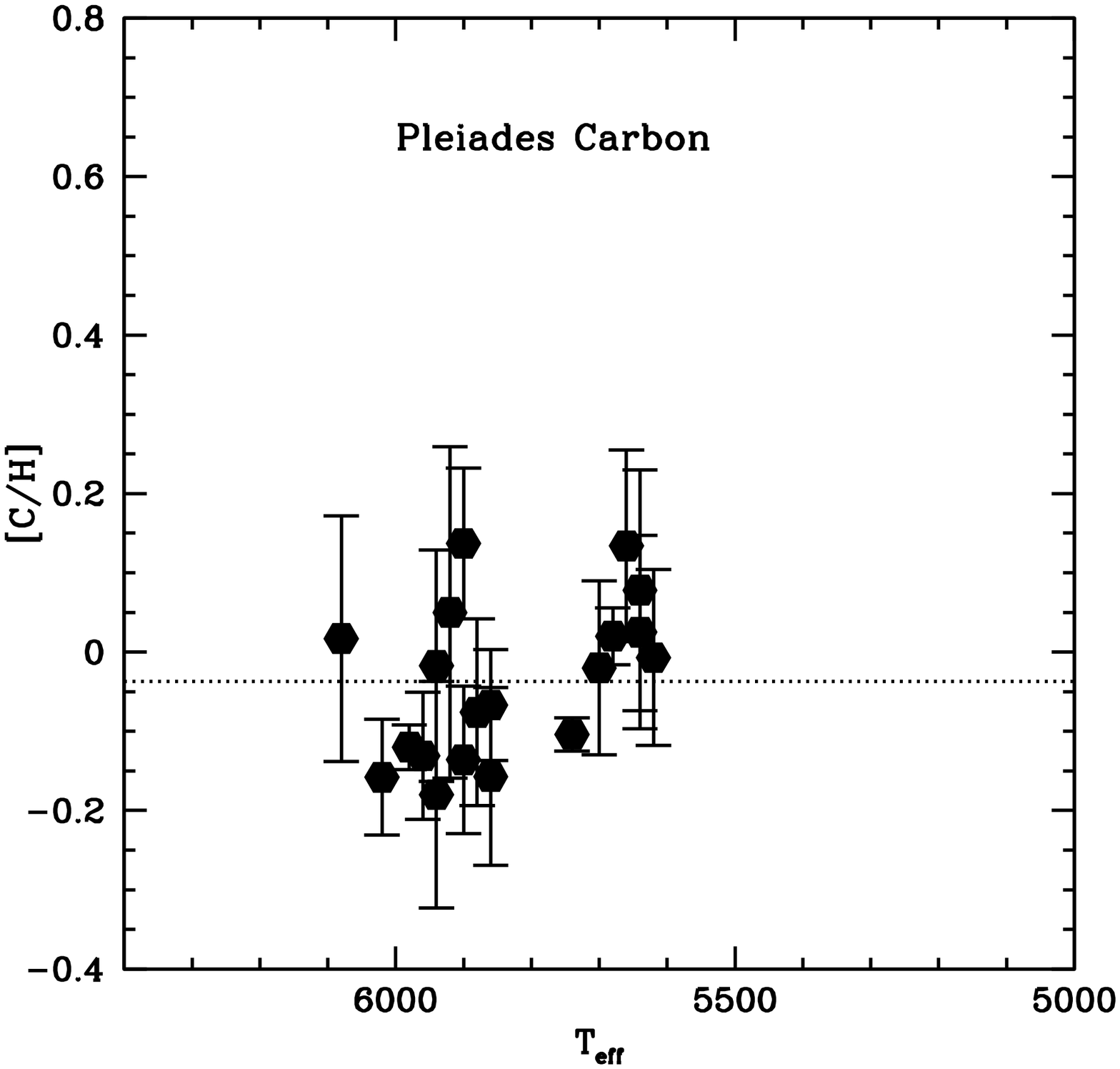}{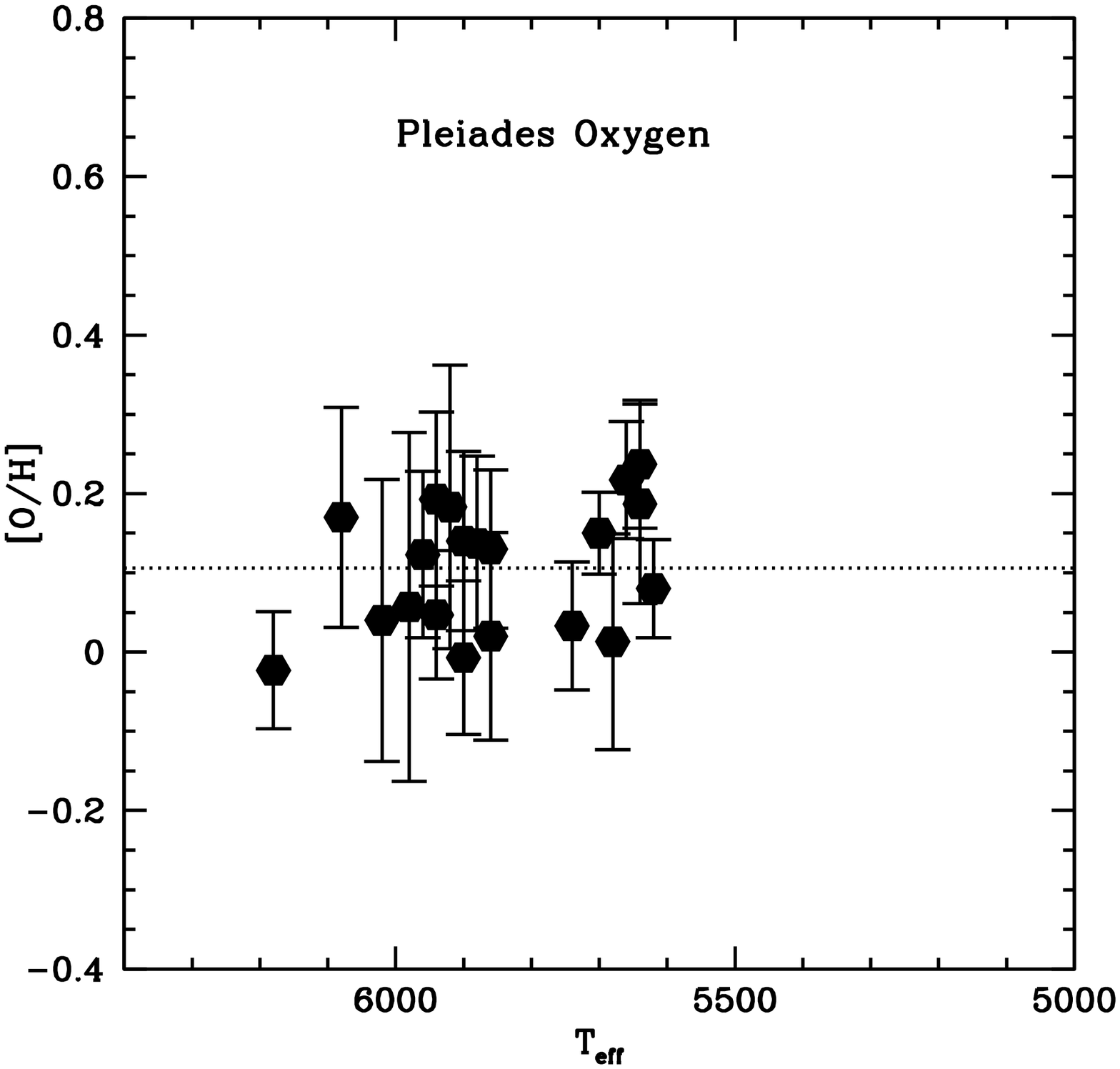}
\caption{Left: Carbon abundances in the Pleiades.  Right: Oxygen abundances in
the Pleiades.  The horizontal dotted lines show the cluster means: [C/H] =
$-$0.04 $\pm$0.02 and [O/H] = +0.11 $\pm$0.02.}
\end{figure}

\subsection{M 67}

The Keck/HIRES spectra of our 14 M 67 stars have S/N ratios ranging from 70 to
140.  The stars we selected were similar to the sun and have a range in
$T_{\rm eff}$ of 5600 - 6000 K.  Figure 14 (left) shows the color-magnitude
diagram for M 67 with our stars as open circles and the cluster represented by
a 5 Gyr isochrone from Vanden Berg (1985).  The [Fe/H] values for the M 67
stars are shown in Figure 14 (right) where the cluster mean is $-$0.05
$\pm$0.01.

\begin{figure}[!ht]
\plottwo{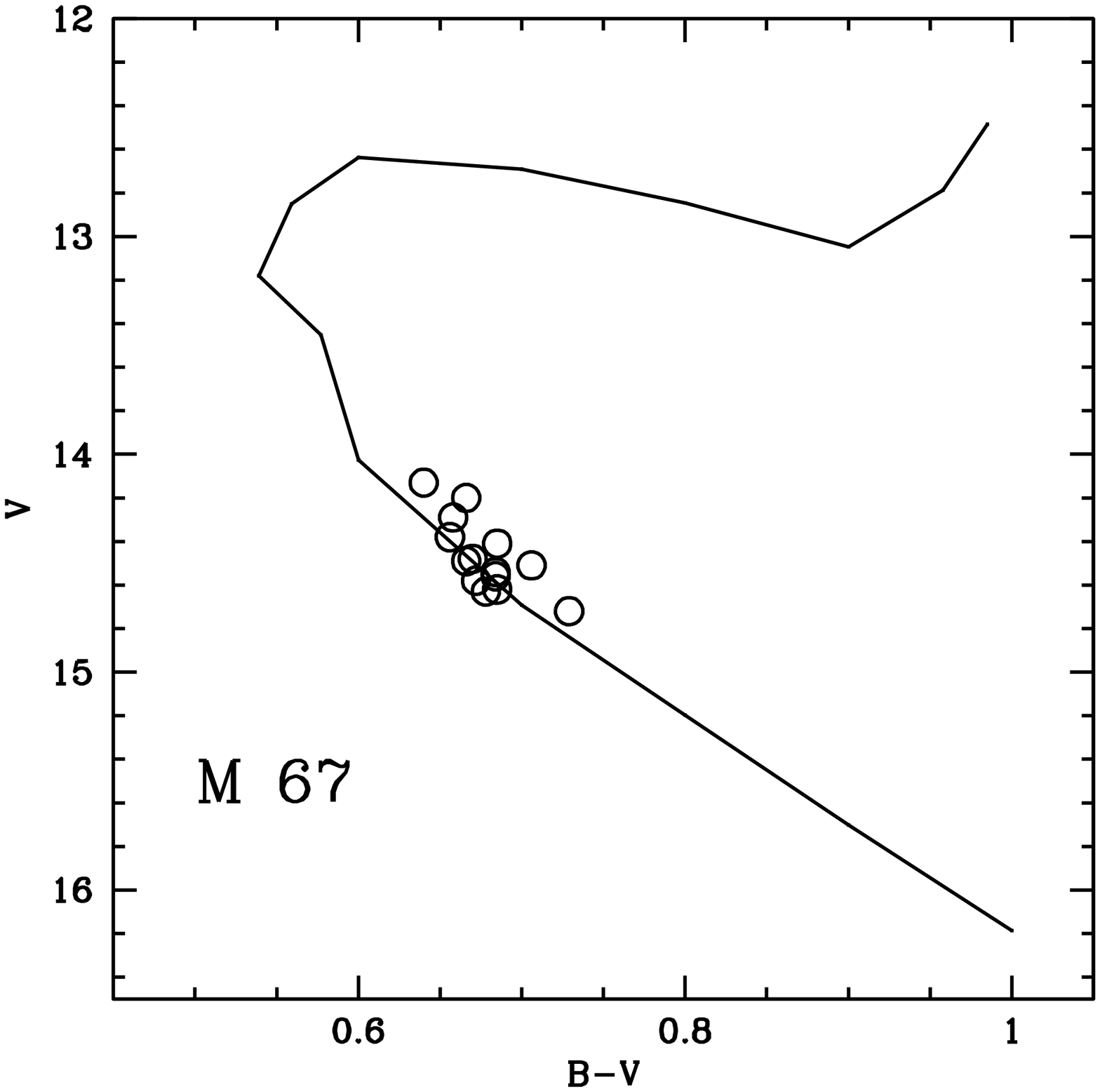}{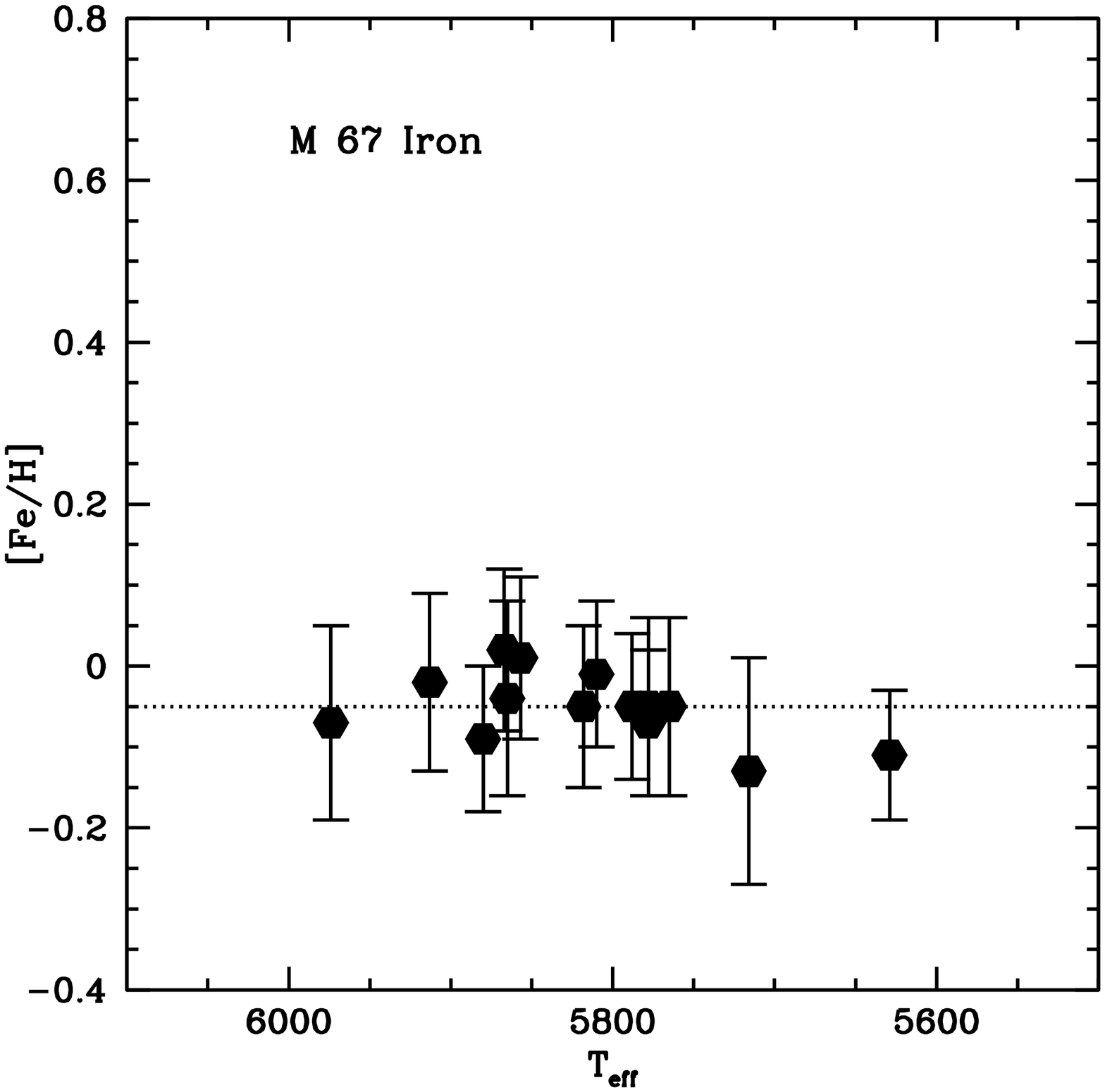}
\caption{Left: The color-magnitude diagram for M 67.  The line is the 5 Gyr
isochrone from Vanden Berg (1985) and the stars we observed are open circles.
Right: The [Fe/H] values for each star plotted against its temperature.  The
cluster mean [Fe/H] = $-$0.05 $\pm$0.01 (dotted line).}
\end{figure}

All of our M 67 stars are cooler than 6200 K (the hottest is 5974 K) and
hotter than 5500 K (the coolest is 5716 K).  The 14 stars are in good
agreement with each other in their values of [C/H] and [O/H].  Those results
are shown in Figure 15.  The cluster means are [C/H] = $-$0.02 $\pm$0.02 and
[O/H] = +0.01 $\pm$0.03.  The abundances of Fe, C, and O are quite close to
solar for the old open cluster, M 67.

\begin{figure}[!ht]
\plottwo{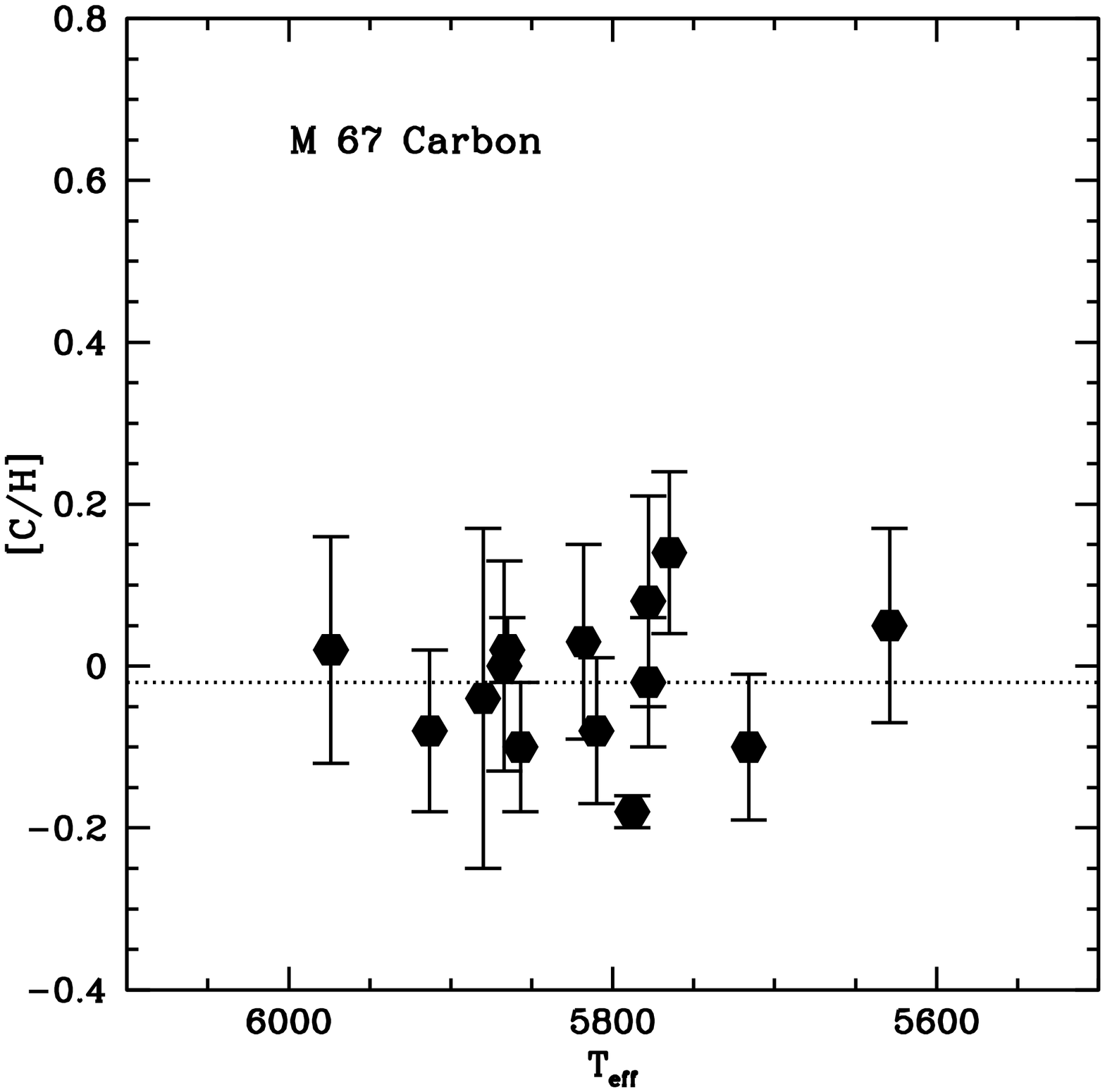}{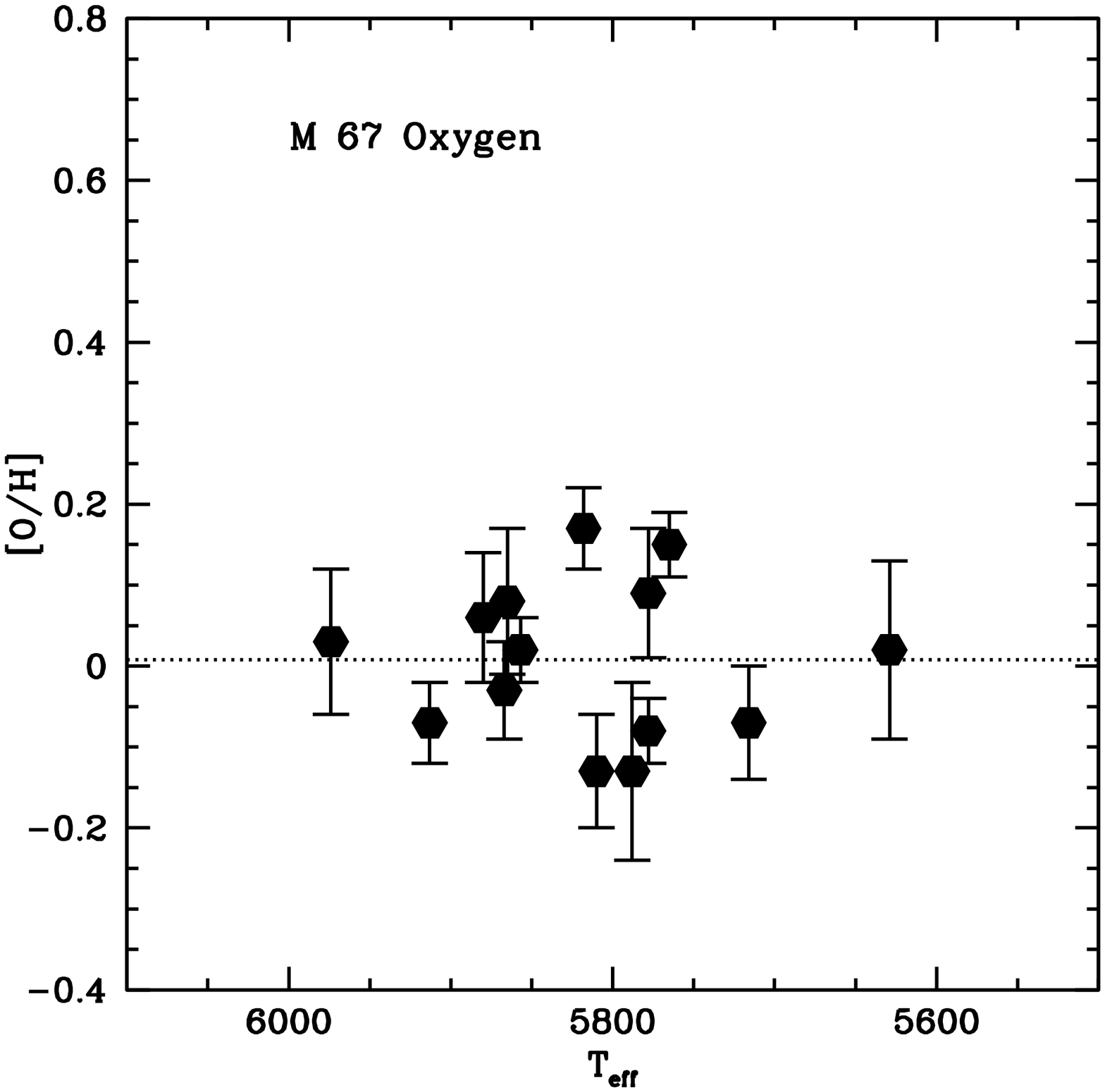}
\caption{Left: Carbon abundances in the M 67.  Right: Oxygen abundances in the
M 67.  The horizontal dotted lines show the cluster means: [C/H] = $-$0.02
$\pm$0.02 and [O/H] = +0.01 $\pm$0.03.}
\end{figure}

\section{Carbon and Oxygen Summary}

The results for the three clusters are summarized in Table 1.  Although there
are only three clusters discussed here, it seems apparent that there is no
evidence of an age-metallicity effect in the disk clusters.  The differences
found in the composition of the clusters could result from the composition in
the local interstellar material from which they were formed, i.e. a ``place of
origin'' effect.  

\begin{table}[!ht]
\caption{Open Cluster Abundances}
\smallskip
\begin{center}
{\footnotesize
\begin{tabular}{lrccccccccc}
\tableline
\noalign{\smallskip}


Cluster & Age & [Fe/H] & $\sigma$
& [C/H] & $\sigma$ & [O/H]
& $\sigma$ & [C/Fe] & [O/Fe] & [O/C] \\
\noalign{\smallskip}
\tableline
\noalign{\smallskip}
Pleiades & 70 Myr  & +0.06   & 0.02 & $-$0.04 & 0.02 & +0.11 & 0.02 & $-$0.10
& +0.05   & +0.15 \\
Hyades	 & 700 Myr & +0.18   & 0.01 & +0.16   & 0.02 & +0.17 & 0.02 & $-$0.02
& $-$0.01 & +0.01 \\
M 67	 & 5 Gyr   & $-$0.05 & 0.01 & $-$0.02 & 0.02 & +0.01 & 0.03 & +0.03
& +0.06   & +0.03 \\
\noalign{\smallskip}
\tableline
\end{tabular}
}
\end{center}
\end{table}

The Hyades is enhanced in Fe, C, and O relative to solar by
about +0.17 dex.  For M 67 the amount of the three elements is similar to that
in the Sun.  Although the Pleiades seems slightly enhanced in Fe and O by
about +0.05 dex compared to the Sun, it seems to be low in C with [C/H] =
$-$0.04 and [C/Fe] = $-$0.09.

\acknowledgements 

I am grateful for the help of several graduate and undergraduate students on
aspects of the yet-unpublished part of this work: M 67 - Elizabeth Barrett and
John Lakatos; Hyades - Jennifer Beard and Edward Lever; Pleiades - Hai Fu.  I
am pleased to acknowledge all the clever help of my various co-authors in the
published papers, especially Jeremy R. King, Eric Armengaud, Constantine
Deliyannis, Alex Stephens, Elizabeth McGrath and David Lambert (to whom this
symposium is dedicated).  I am happy to thank Megan Novicki for important
contributions to the presentation.  Figures 1-6 are reproduced by permission
of the American Astronomical Society.  The work reported here has been
supported by N.S.F. grants AST 00-97945 and AST 00-97955 and by NASA/STScI
grants HST-GO-08770 and HST-GO-09048.

\end{document}